\documentclass[conference]{IEEEtran}
\IEEEoverridecommandlockouts
\usepackage{cite}
\usepackage[hidelinks]{hyperref}
\usepackage{graphicx}
\usepackage{subfigure}
\usepackage{textcomp}
\usepackage{xcolor}
\usepackage[norelsize,ruled,vlined,linesnumbered]{algorithm2e}
\usepackage{algorithmic}
\usepackage{amsthm}
\usepackage{amsmath,amssymb,amsfonts}
\usepackage{booktabs}
\usepackage{balance}
\usepackage{url}

\newtheoremstyle{mystyle}
    {2.0mm}
    {2.0mm}
    {\it}
    {0.0mm}
    {\scshape}
    {.}
    { }
    {}
\theoremstyle{mystyle}
\newtheorem{definition}{Definition}
\newtheorem{example}{Example}

\newtheorem{theorem}{Theorem}
\newtheorem{corollary}{Corollary}
\newtheorem{remark}{Remark}
\newtheorem{problem}{Problem}

\newcommand{\vs}{\vspace{1.5mm}}
\newcommand{\wsq}{\hspace{\fill}$\square$}
\newcommand{\hint}{HINT$^{m} \,$}
\SetKwComment{Comment}{$\triangleright$\ }{}

\def\BibTeX{{\rm B\kern-.05em{\sc i\kern-.025em b}\kern-.08em
    T\kern-.1667em\lower.7ex\hbox{E}\kern-.125emX}}
\begin{document}

\title{Independent Range Sampling on Interval Data (Longer Version)}

\author{
\IEEEauthorblockN{Daichi Amagata}
\IEEEauthorblockA{\textit{Graduate School of Information Science and Technology} \\
\textit{Osaka University}\\
Japan \\
amagata.daichi@ist.osaka-u.ac.jp}
}

\maketitle

\begin{abstract}
Many applications require efficient management of large sets of intervals because many objects are associated with intervals (e.g., time and price intervals).
In such interval management systems, range search is a primitive operator for retrieving and analysis tasks.
As dataset sizes are growing nowadays, range search results are also becoming larger, which may overwhelm users and incur long computation time.
Because applications are usually satisfied with a subset of the result set, it is desirable to efficiently obtain only small samples from the result set.
We therefore address the problem of independent range sampling on interval data, which outputs $s$ random samples that overlap a given query interval and are independent of the samples of all previous queries.
To efficiently solve this problem theoretically and practically, we propose a variant of an interval tree, namely the augmented interval tree (or AIT), and we show that there exists an exact algorithm that needs $O(n\log n)$ space and $O(\log^{2}n + s)$ time, where $n$ is the dataset size.
The simple structure of an AIT provides flexible extensions: (i) its time and space complexities respectively become $O(\log^{2}n + s)$ expected and $O(n)$ by bucketing intervals and (ii) it can deal with weighted intervals and outputs $s$ weighted random samples in $O(\log^{2}n + s\log n)$ time.
We conduct extensive experiments on real datasets, and the results demonstrate that our algorithms significantly outperform competitors.
\end{abstract}


\section{Introduction}  \label{sec:introduction}
An interval $x$ is defined as a pair of one-dimensional points and $x = [x.l, x.r]$ ($x.l \leq x.r$), where $x.l$ and $x.r$ respectively are the left- and right-endpoints of $x$.
Many applications require efficient interval data management.
For example, in temporal databases, each temporal object is a pair of starting and ending times, and the interval represents active time \cite{nishio2020lamps}.
Uncertain databases also employ similar data management, because some attribute values of uncertain objects are not fixed, and they are represented as bounding intervals \cite{zhang2010efficient}.
Moreover, stock and cryptocurrency applications record fluctuations of their prices, so the prices of stocks and cryptocurrency are represented by intervals (e.g., $[\min, \max]$) based on a time unit.

One of the most important operators for interval retrieval and analysis is range query \cite{christodoulou2022hint,christodoulou2023hint,behrend2019period}.
Consider a set $X$ of $n$ intervals, and given a query interval $q = [q.l, q.r]$, the range search problem finds a set of all intervals in $X$ overlapping $q$.
This set is represented as $q \,\cap X = \{x \,|\, x \in X, q.l \leq x.r \wedge x.l \leq q.r\}$.
Some examples that use range queries are as follows.
\begin{description}
    \setlength{\leftskip}{-3.0mm}
    \item[Ex. 1.]   Vehicle (e.g., taxi) management systems:
                    Show vehicles that were active between 17:00 and 22:00 a week ago.
    \item[Ex. 2.]   Book management systems or libraries:
                    Collect books that were sold or borrowed in the last month.
    \item[Ex. 3.]   Historical cryptocurrency databases:
                    Show when the price of Bitcoin (BTC) falls in [30,000 40,000] dollars.
\end{description}
The search results of the above queries are easily handled if their sizes are small.
If the sizes are large, however, users would be overwhelmed, and analysis tasks may become difficult.
In addition, the range search time becomes longer, because range search algorithms incur $\Omega(|q \,\cap\ X|)$ time.
Dataset sizes are growing nowadays, and $|q \cap X| = \Omega(n)$ in practice, so the output sizes of range queries also increase.
Then, simply reporting all intervals overlapping $q$ can yield difficulty in subsequent analysis tasks and long running times.
Assume, for example, that a taxi management system wants to visualize active taxis to see their location distribution through a range query $q$ as in Ex. 1.
If $|q \cap X|$ is huge (e.g., more than hundreds of thousands), such visualization incurs a long delay \cite{xie2021spatial}.
In addition, suppose that we are interested in shopping statistics for each month from 2018 to 2023 on an e-commerce site (e.g., an online bookstore as in Ex. 2) to investigate whether we have some purchase pattern changes before, during, and after the COVID-19 pandemic.
It is easy to imagine that there are a significant number of transactions\footnote{Each transaction is considered as an interval, where its left-endpoint is the time when the transaction started (e.g., the item was purchased online) and its right-endpoint is the time when the transaction was completed (e.g., the time when the item was delivered).} in each month, so collecting them incurs a substantial computational cost.

\vs
\noindent
\textbf{Motivation.}
Fortunately, many applications do not require complete result sets but are satisfied with their approximate results (e.g., small subsets) to enable interactive analysis \cite{wang2015spatial,amagata2022learned}.
More specifically, random samples from the result sets are sufficient \cite{simpler2023aoyama,har2019near,aumuller2022sampling,aumuller2020fair,olken1995random,tao2022algorithmic,xie2021spatial,afshani2019independent,hu2014independent,olken1989random,afshani2017independent}.
In the above taxi visualization and e-commerce applications, random samples are sufficient to see the distribution and estimate statistics.
Also, in cryptocurrency databases, the price fluctuation is usually recorded in a fine-grained manner (e.g., every few minutes).
As for Ex. 3, using random samples can reduce redundant data while retaining necessary intervals (to know when).
It is also important to notice that retrieving only random samples can significantly reduce query processing time.
Then, we see that range \textit{sampling} queries solve the drawbacks of range queries.

Motivated by the above observations, this paper addresses the problem of \textit{independent range sampling (IRS) on interval data}, which, given a query interval $q$, retrieves $s$ random samples from $q \cap X$, and the $s$ samples must be independent of all previous query results.
(Its formal definition appears in Section \ref{sec:preliminary:problem-definition}.)
This paper is the first work that addresses this problem.
Note that independence is important, particularly when some statistical property is required \cite{tao2022algorithmic}, like the above sampling examples.
Without independence, biased results may be observed, leading to wrong conclusions \cite{xie2021spatial}.

\vs
\noindent
\textbf{Challenge.}
As the above applications require only $s$ random samples, the running times of IRS queries should be sensitive only to $s$.
That is, the time complexity (and the space complexity) of an IRS algorithm should be $\tilde{O}(s)$, where $\tilde{O}(\cdot)$ hides any polylogarithmic factors.
This is a natural requirement of applications that specify output sizes for practical and theoretical efficiency \cite{gan2017dynamic}.

A straightforward approach to solving the IRS problem on interval data is to employ a state-of-the-art range search algorithm (e.g., \hint \cite{christodoulou2022hint}) to obtain $q \cap X$ and then sample $s$ random intervals from $q \cap X$.
Although this approach correctly solves the problem, the merit of range sampling is lost, since $\Omega(|q \cap X|) = \Omega(n)$ time is required.
One may come up with an idea of using the IRS algorithm for one-dimensional data \cite{hu2014independent}.
This algorithm stores one-dimensional points in a sorted array and uses two binary searches to find the boundary indices that identify the borders covered by a given query range.
Then, simple random sampling from the border range achieves an equal and independent sampling probability.
One may consider that this algorithm can solve our problem by using the left- and right-endpoints of intervals, but this cannot solve our problem correctly.
For each of the intervals fully covered by the query range, both its left- and right-endpoints are contained in the range.
On the other hand, for each of the intervals partially covered by the query range, either its left- or right-endpoint is contained in the range.
This means that, to enable equal sampling probability for each interval overlapping the query range, we have to identify whether it is fully covered or not.
That is, we need to access all intervals in the range in the worst case, which incurs $O(n)$ time, as with search-based algorithms.
Another approach is to prepare samples offline \cite{wang2015spatial}, but this trivially violates independence.

In \cite{xie2021spatial}, several IRS algorithms for spatial points were proposed.
They can solve our problem because intervals can be mapped into a 2-dimensional space \cite{rahul2014general}.
However, even the best algorithm among them returns $s$ random samples in $O(\sqrt{n} + s)$ expected time, which is still sensitive to the dataset size.
(In addition, the other algorithms incur $O(n)$ time, which loses the efficiency of random sampling.)
To summarize, the existing techniques (including IRS algorithms for one-dimensional and spatial data) cannot efficiently and/or correctly solve the IRS problem on interval data, and achieving an $\tilde{O}(s)$ time algorithm is not trivial.

\vs
\noindent
\textbf{Contribution.}
We overcome the above challenge and propose several algorithms that run in $\tilde{O}(s)$ time without losing independence and equal sampling probability.
We summarize our contributions below\footnote{This paper is a longer version of \cite{amagata2024independent}.}.

\vs
\noindent
$\bullet \,$ \underline{\textit{AIT: Augmented Interval Tree} (Section \ref{sec:uniform}).}
We propose AIT, which augments the most famous data structure for interval data, namely Edelsbrunner’s interval tree \cite{edelsbrunner1980dynamic}.
It is important to note that the original interval tree structure supports efficient processing of stabbing queries but \textit{does not support efficient range-related operators}, e.g., range queries, because $O(n)$ time is incurred in the worst case.
Our augmentation overcomes this drawback and derives a new technique that can identify the space where \textit{only} $q \cap X$ exists in $\tilde{O}(1)$ time.
Thanks to this technique, we can randomly sample $s$ intervals while preserving the independence, resulting in $\tilde{O}(s)$ time.
As a side product, the AIT structure supports a range \textit{counting} query, which returns $|q \cap X|$, in $\tilde{O}(1)$ time.
Furthermore, we show how to efficiently support interval insertions and deletions.

\vs
\noindent
$\bullet \,$ \underline{\textit{AIT-V: AIT with virtual intervals} (Section \ref{sec:uniform:aitv}).}
Because an AIT needs $O(n\log n)$ space, we next consider achieving $O(n)$ space complexity.
This can be done by allowing expected time.
We bundle intervals to make ``virtual'' intervals, and these virtual intervals are maintained by an AIT.
We show that this black-box bundling approach (with size constraint) enables $\tilde{O}(s)$ expected time and $O(n)$ space.

\vs
\noindent
$\bullet \,$ \underline{\textit{AWIT: Augmented Weighted Interval Tree} (Section \ref{sec:weighted}).}
As the second extension of the AIT, we propose AWIT to deal with weighted intervals.
This weighted case is also an important setting and has been studied in existing works (with different data formats) \cite{afshani2017independent,afshani2019independent,xie2021spatial,zhange2023efficient}.
If each interval $x_{i}$ has a weight $w(x_{i})$, its sampling probability should be $\frac{w(x_{i})}{\sum_{x_{j} \in q \cap X} w(x_{j})}$.
Although achieving this for arbitrary queries is not trivial without accessing all intervals in $q \cap X$, we demonstrate that the AWIT structure keeps $\tilde{O}(s)$ time in this weighted case.

\vs
\noindent
$\bullet \,$ \underline{\textit{Empirical evaluation.} (Section \ref{sec:experiment})}
We conduct extensive experiments on real-world datasets.
The results demonstrate that our algorithms are usually one order of magnitude faster than competitors in non-weighted and weighted interval cases.

\begin{table}[!t]
    \centering
    \caption{Comparison of time and space complexities, where $n$, $q \cap X$, and $s$ are the dataset size, range search result set, and sample size, respectively.
    Note that $O(\cdot)$* suggests expectation.}
    \label{tab:complexity}
    \begin{tabular}{lllc} \toprule
                                            & Time                      & Space         & Weighted?     \\ \midrule
        \hint \cite{christodoulou2022hint}  & $\Omega(|q \cap X|)$      & $O(n)$        & \checkmark    \\
        KDS \cite{xie2021spatial}           & $O(\sqrt{n} + s)$*        & $O(n)$        &               \\
        KDS (weighted)                      & $O(\sqrt{n} + s\log n)$*  & $O(n)$        & \checkmark    \\ \hline
        AIT                                 & $O(\log^{2}n + s)$        & $O(n\log n)$  &               \\
        AIT-V                               & $O(\log^{2}n + s)$*       & $O(n)$        &               \\
        AWIT                                & $O(\log^{2}n + s\log n)$  & $O(n\log n)$  & \checkmark    \\ \bottomrule
    \end{tabular}
\end{table}

\vs
Table \ref{tab:complexity} compares our theoretical contribution with the existing ones.
Since $|q \,\cap X| = \Omega(n)$ in practice, our algorithms are faster than the range search algorithms.
In addition, for sufficiently large $n$, we have $\log^{2} n < \sqrt{n}$, so our algorithms are faster than KDS.
These results clarify the superiority and novelty of our algorithms.

\vs
\noindent
\textbf{Organization.}
The rest of this paper is organized as follows.
Section \ref{sec:preliminary} introduces preliminary information to present our algorithms.
Section \ref{sec:uniform} presents the AIT and our algorithms for non-weighted intervals, and, in Section \ref{sec:weighted}, we extend the AIT for weighted intervals.
We report our experimental results in Section \ref{sec:experiment}, and Section \ref{sec:related_work} reviews related works.
We conclude this paper in Section \ref{sec:conclusion}.

\section{Preliminary}   \label{sec:preliminary}

\subsection{Problem Definition}    \label{sec:preliminary:problem-definition}
Let $X$ be a set of $n$ intervals.
We use $x$ to denote an interval in $X$, and $x$ is defined as a pair of its left-endpoint $x.l$ and right-endpoint $x.r$, i.e., $x = [x.l, x.r]$ ($x.l \leq x.r$).
Let $x \cap x'$ mean that $x$ overlaps $x'$, i.e., $(x.l \leq x'.r) \wedge (x'.l \leq x.r)$.
Similarly, we define $x \cap X$ as $\{x' \,|\, x' \in X, x \cap x'\}$.
Now we are ready to formally define the problem of independent range sampling (IRS) on interval data.

\begin{problem}[\textsc{IRS on interval data}]  \label{problem:uniform}
Given an interval set $X$, a query interval $q$, and a sample size $s$, this problem returns a set $S$ of $s$ intervals, each of which is picked from $q \cap X$ uniformly at random.
\end{problem}

\noindent
The above definition means that the returned $s$ random samples are independent of the outputs of all previous IRS queries.

Next, if each interval $x \in X$ has a weight $w(x)$, its sampling probability should be proportional to $w(x)$.
The IRS problem on weighted interval data is formally defined as follows:

\begin{problem}[\textsc{IRS on weighted interval data}] \label{problem:weighted}
Given an interval set $X$, a query interval $q$, and a sample size $s$, this problem returns a set $S$ of $s$ weighted intervals randomly picked from $q \cap X$.
The sampling probability of $x \in q \cap X$ is $\frac{w(x)}{\sum_{x_{i} \in q \cap X} w(x_{i})}$.
\end{problem}

We assume that $X$ is memory resident, as with the existing works that deal with interval data \cite{christodoulou2022hint,christodoulou2023hint,bouros2017forward,rahul2014general,piatov2016interval,bouros2021memory,cafagna2017disjoint,piatov2021cache,dignos2014overlap}, because recent machines equip memory with large capacity.
The objective of this paper is to design theoretically and practically efficient algorithms for the IRS problems on non-weighted and weighted interval data without violating the independent sampling probability.

\subsection{Interval tree}  \label{sec:preliminary:interval-tree}
We introduce Edelsbrunner’s interval tree \cite{edelsbrunner1980dynamic} since we use it as a building block of our data structure.
Henceforth, we call it the interval tree simply.
Its structure is essentially similar to the binary tree structure.
Each of its nodes $u_{i}$ consists of the following components:
\begin{itemize}
    \setlength{\leftskip}{-2.5mm}
    \item   $c_{i}$: the central point.
    \item   $L^{l}_{i}$: a list containing all intervals $x$ such that $x.l \leq c_{i} \leq x.r$.
            The intervals in $L^{l}_{i}$ are sorted in ascending order of the left-endpoint.
    \item   $L^{r}_{i}$: a list containing the same intervals in $L^{l}_{i}$.
            The intervals in $L^{r}_{i}$ are sorted in ascending order of the right-endpoint.
    \item   $u^{l}_{i}$: the left child node of $u_{i}$.
            The intervals $x'$ maintained by the sub-tree rooted at $u^{l}_{i}$ satisfy $x'.r < c_{i}$.
    \item   $u^{r}_{i}$: the right child node of $u_{i}$.
            The intervals $x'$ maintained by the sub-tree rooted at $u^{r}_{i}$ satisfy $x'.l > c_{i}$.
\end{itemize}
We use $L^{l}_{i}[j]$ ($L^{r}_{i}[j]$) to denote the $j$-th interval in $L^{l}_{i}$ ($L^{r}_{i}$).
This paper assumes that the index starts from 1.

\vs
\noindent
\textbf{Construction.}
Given a set $X$ of intervals, we first create a root node $u_{root}$.
We compute the central (typically median) point among all left- and right-endpoints in $X$.
We then compute all intervals $x$ such that $x.l \leq c_{root} \leq x.r$ to make $L^{l}_{root}$ and $L^{r}_{root}$.
At the same time, we obtain sets $X_{l} = \{x \,|\, x \in X,  x.r < c_{root}\}$ and $X_{r} = \{x \,|\, x \in X,  x.l > c_{root}\}$.
We create a left (right) child node of $u_{root}$ from $X_{l}$ ($X_{r}$).
The interval tree of $X$ is built in this recursive manner until a given node has $X_{l} = X_{r} = \varnothing$.

\begin{remark}  \label{remark:interval-tree}
The space complexity and the height of an interval tree are $O(n)$ and $O(\log n)$, respectively.
The interval tree structure supports efficient stabbing query processing, and a stabbing query is a special case of the range query $q$ such that $q.l = q.r$.
It can be processed in $O(\log n + K)$ time, where $K$ is the output size.
However, this structure does not support efficient range query processing.
This is trivial: if a query interval $q$ covers $X$, we need to traverse all nodes, deriving $O(n)$ worst time.
\end{remark}

\noindent
This remark suggests that it is not trivial to identify the space where only $q \cap X$ exists in $\tilde{O}(1)$ time by using the interval tree structure.

\subsection{Weighted Sampling Methods}  \label{sec:preliminary:sampling}
Last, we introduce existing weighted sampling methods because our algorithms employ them.

\vs
\noindent
\textbf{Walker's alias method \cite{walker1974new}.}
Given $n$ objects $o_{1}$, ..., $o_{n}$ associated with weights $w_{1}, ..., w_{n}$, this method needs $O(n)$ pre-processing time to build a data structure called \textit{alias} with $O(n)$ space.
This alias achieves $O(1)$ time weighted sampling, and the sampling probability of $w_i$ is $\frac{w_i}{\sum_{j=1}^{n}w_{j}}$.

\vs
\noindent
\textit{Alias building.}
The alias consists of $n$ cells.
The constraints of the alias structure are that (i) each cell can have at most two objects and (ii) the $n$ cells have the pre-defined weight capacity $\tau$, which is based on $\sum_{j=1}^{n}w_{j}$.
The objects with weights not larger than $\tau$ are first assigned to distinct cells, and each of these cells maintains $\langle i,w_{i}\rangle$ ($w_{i} \leq \tau$).
Then, the remaining objects are assigned to empty cells and/or cells having $\langle i,w_{i}\rangle$ where $w_{i} < \tau$.
Specifically, $o_{j}$ with $w_{j} > \tau$ is maintained by multiple cells, and $w_j$ is distributed to these cells according to their remaining capacities\footnote{If a given cell is empty, it maintains $\langle j,\tau\rangle$.
On the other hand, if it already has $\langle i,w_{i}\rangle$, it additionally maintains $\langle j,\tau - w_{i}\rangle$.
The cells maintaining $o_j$ have one of these cases, and the sum of their corresponding weights is $w_{j}$.}.

\vs
\noindent
\textit{Sampling.}
When sampling an object, we first pick one cell uniformly at random.
If this cell maintains a single object, we return it as a sample.
Otherwise, we generate a random weight in $[0,\tau]$ and pick one of the two objects maintained by the cell based on the generated and their weights.

\vs
\noindent
\textbf{Cumulative sum method.}
Given $n$ weighted objects, this method needs $O(n)$ pre-processing time to build an array $A$ with $n$ elements and consumes $O(n)$ space.
From this array, we can pick $o_{i}$ in $O(\log n)$ time with probability $\frac{w_i}{\sum_{j=1}^{n}w_{j}}$.

\vs
\noindent
\textit{Array building.}
Each element of $A$ maintains the cumulative sum of the weights.
Specifically, $A[j]$ maintains $\sum_{i=1}^{j}w_{i}$, so building $A$ is straightforward.

\vs
\noindent
\textit{Sampling.}
We generate a random weight $w$ in $[0,\sum_{j=1}^{n}w_{j}]$.
If $A[k-1] < w \leq A[k]$, we return $k$ (i.e., $o_{k}$) as a sample.
Finding $k$ can be done by a binary search, so $O(\log n)$ time is required for one-time sampling.
Notice that $k$ can be returned iff we have $\sum_{j=1}^{k-1}w_{j} < w \leq \sum_{j=1}^{k}w_{j}$, and its probability is $(\sum_{j=1}^{k}w_{j} - \sum_{j=1}^{k-1}w_{j}) / \sum_{j=1}^{n}w_{j} = w_k / \sum_{j=1}^{n}w_{j}$.

\section{Non-weighted Interval Case}  \label{sec:uniform}
This section solves Problem \ref{problem:uniform} and proposes $\tilde{O}(s)$ time algorithms.

\vs
\noindent
\textbf{Main idea.}
Our first idea for efficiently solving Problem \ref{problem:uniform} is to exploit the interval tree structure, particularly the sorted lists of each node.
Consider node $u_i$ of an interval tree.
The intervals overlapping the central point $c_i$ are maintained by $L^{l}_{i}$ and $L^{r}_{i}$.
Given a query interval $q$, the intervals in $L^{l}_{i}$ (or $L^{r}_{i}$) overlapping $q$ appear in a sequential manner, since they are sorted.
To obtain random samples, it is sufficient to know only the pair of the left and right indices of $L^{l}_{i}$ (or $L^{r}_{i}$) that represent an ``overlapping range'' in this node.
This can be done in $O(\log n)$ time with a single binary search, because $|L^{l}_{i}| = O(n)$.

\begin{figure}[!t]
    \centering
    \includegraphics[width=0.65\linewidth]{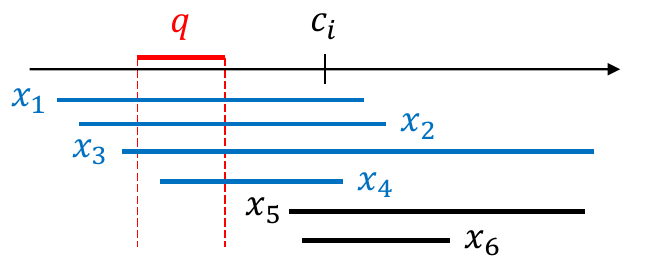}
    \caption{Illustration of our main idea.
    Node $u_i$ of an interval tree maintains $\{x_1, x_2, x_3, x_4, x_5, x_6\}$, and $L^{l}_{i} = [x_1, x_2, x_3, x_4, x_5, x_6]$.
    A query $q$ overlaps $x_1$, $x_2$, $x_3$, and $x_4$.}
    \label{fig:idea}
\end{figure}

\begin{example} \label{example:idea}
Fig. \ref{fig:idea} depicts node $u_i$ of an interval tree.
It maintains $\{x_1, x_2, x_3, x_4, x_5, x_6\}$, and $L^{l}_{i} = [x_1, x_2, x_3, x_4, x_5, x_6]$.
Due to the interval tree structure, $x_{j}.l \leq c_i \leq x_{j}.r$ ($j \in [1,6]$).
Then, a binary search on $L^{l}_{i}$ with a query point $q.r$ can find the boundary such that $L^{l}_{i}[j].l \leq q.r < L^{l}_{i}[j+1].l$ ($j = 4$ in this example).
Consequently, we see that $q \cap L^{l}_{i}[j]$ ($j \in [1,4]$) without enumerating the intervals overlapping $q$.
\end{example}

Doing the above operation for each node (which is not pruned) obtains a set of pairs of left and right indices, and this set represents the space where only $q \cap X$ exists.
However, we need to traverse $O(n)$ nodes in the worst case when we use the interval tree, see Remark \ref{remark:interval-tree}.
Our second idea is, hence, to bound the number of traversed nodes with the height of an interval tree, i.e., $O(\log n)$.
Then, only $O(\log n) \times O(\log n) = \tilde{O}(1)$ time is required to run binary searches for identifying the space where only $q \cap X$ exists.
To obtain this desirable time without losing correctness, we augment the interval tree structure.
\textit{Finding the space where only $q \cap X$ exists in $\tilde{O}(1)$ time with at most a single binary search for each accessed node is a new result and the main difference} to the existing techniques that can be employed for our problem.

\vs
\noindent
\textbf{Overview.}
We build our data structure in the pre-processing phase, and our online algorithm consists of two phases.
The first phase collects a set of nodes (of our data structure) maintaining the intervals overlapping $q$ in the way described above.
The second phase randomly samples $s$ intervals from the data structure obtained in the first phase.

\subsection{AIT: Augmented Interval Tree}   \label{sec:uniform:ait}
Assume that $X$ is indexed by an interval tree.
Given a query interval $q$, not to miss any intervals that overlap $q$, we need to access all nodes that possibly maintain such intervals.
Then, assume that we now access node $u_{i}$ such that $q.l \leq c_{i} \leq q.r$.
In this case, we have to traverse both its left and right child nodes, because the sub-trees rooted at these nodes may maintain intervals overlapping $q$.
The interval tree structure cannot bound the number of nodes facing this case, resulting in $O(n)$ time to deal with a range query.
This issue is common in other state-of-the-art data structures for interval data, e.g., \hint \cite{christodoulou2022hint}.
We eliminate this drawback by extending the interval tree structure, and our structure enables us to face the above case \textit{at most once}.

\vs
\subsubsection{\textbf{Structure}}  \label{sec:uniform:ait:structure}
\begin{figure}
    \centering
    \includegraphics[width=0.95\linewidth]{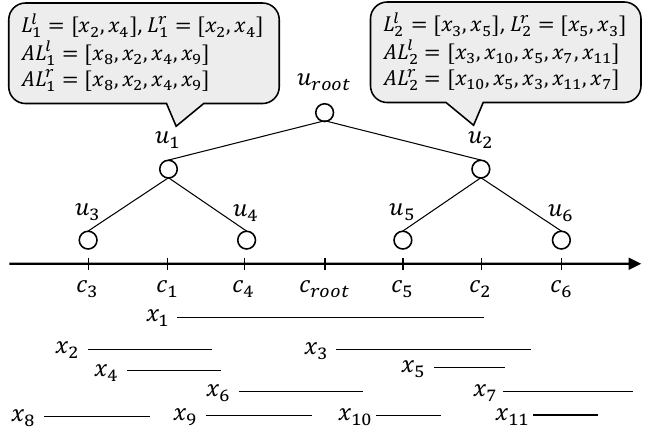}
    \caption{Illustration of an AIT on a small dataset set $X = \{x_1, x_2, ..., x_{11}\}$.
    The top part depicts the AIT on $X$, whereas the bottom part shows $x_{i}$ ($i \in [1,11]$).}
    \label{fig:ait}
\end{figure}

We propose AIT (Augmented Interval Tree), a variant of the interval tree, and we augment the node structure of the interval tree.
Our augmentation is the key factor for achieving a non-trivial IRS algorithm that has a solid theoretical performance guarantee and avoids the drawback held by the existing structures.
Specifically, in addition to the components introduced in Section \ref{sec:preliminary:interval-tree}, each node $u_{i}$ of an AIT maintains the following components:
\begin{itemize}
    \setlength{\leftskip}{-3.0mm}
    \item   $AL^{l}_{i}$: a list containing all intervals maintained by the sub-tree rooted at $u_{i}$.
            The intervals in $AL^{l}_{i}$ are sorted in ascending order of the left-endpoint.
    \item   $AL^{r}_{i}$: a list containing the same intervals in $AL^{l}_{i}$.
            The intervals in $AL^{r}_{i}$ are sorted in ascending order of the right-endpoint.
\end{itemize}
As with $L^{l}_{i}$ and $L^{r}_{i}$, we use $AL^{l}_{i}[j]$ ($AL^{r}_{i}[j]$) to denote the $j$-th interval in $AL^{l}_{i}$ ($AL^{r}_{i}$).
An example of an AIT and its space complexity are as follows:

\begin{example} \label{example:ait}
Fig. \ref{fig:ait} illustrates an AIT on a small dataset.
Focus on node $u_{2}$, and $L^{l}_{2}$, $L^{r}_{2}$, $AL^{l}_{2}$, and $AL^{r}_{2}$ are depicted.
The sub-tree rooted at $u_{2}$ has $x_{3}$, $x_{5}$, $x_{7}$, $x_{10}$, and $x_{11}$ (their left-endpoints are larger than $c_{root}$), so they are contained in $AL^{l}_{2}$ and $AL^{r}_{2}$.
\end{example}

\vspace{-2.0mm}
\begin{theorem}[\textsc{Space complexity of AIT}]   \label{theorem:ait:space}
The space complexity of an AIT is $O(n\log n)$.
\end{theorem}

\noindent
\textsc{Proof.}
The height of an interval tree is $O(\log n)$, so the height of an AIT is also $O(\log n)$.
This fact and the definition of $AL^{l}$ ($AL^{r}$) derive that each interval appears in at most $O(\log n)$ nodes.
Then, this theorem is clear.    \wsq

\vspace{2.0mm}
\subsubsection{\textbf{Construction}}
As with the interval tree construction, an AIT is built in a top-down manner and in a pre-processing phase.
The procedure is almost the same as that in Section \ref{sec:preliminary:interval-tree}.
The only difference is that we create $AL^{l}$ and $AL^{r}$.
When we are given $X' \subseteq X$ and create a new node $u_{i}$, all intervals in $X'$ are included in $AL^{l}_{i}$ and $AL^{r}_{i}$.
Note that as long as $X$ is static, the AIT is built only once and accepts arbitrary query intervals and sample sizes.

\subsection{IRS Algorithm for Non-weighted Intervals}   \label{sec:uniform:algorithm}

\subsubsection{\textbf{Observation}} \label{sec:uniform:algorithn:observation}
To start with, we elaborate on the observations about, in the AIT, (i) which sorted list should be used and (ii) which nodes should be traversed.
Given a query interval $q$ and node $u_{i}$ of the AIT, this node has one of the following cases:

\vs
\noindent
\underline{\textit{Case 1: $q.r < c_{i}$.}}
In this case, the query interval is located to the left of $c_{i}$, as shown in Fig. \ref{fig:idea}.
For each $x' \in L^{l}_{i}$ such that $x'.l \leq q.r$, we have $x' \cap q$, because $q.r < c_{i} \leq x'.r$.
From the (augmented) interval tree structure, every interval $x$ maintained in the sub-tree rooted at the right child node of $u_{i}$ has $x.l > c_{i}$.
We can hence prune this sub-tree, and its left child node $u_{i}^{l}$ has to be traversed.

\vs
\noindent
\underline{\textit{Case 2: $c_{i} < q.l$.}}
The query interval is located to the right of $c_{i}$, and for each $x' \in L^{r}_{i}$ such that $q.l \leq x'.r$, we have $x' \cap q$ because $x'.l \leq c_{i} < q.l$.
Every interval $x$ maintained in the sub-tree rooted at the left child node of $u_{i}$ has $x.r < c_{i}$.
This sub-tree can be pruned, and its right child node $u_{i}^{r}$ has to be traversed.

\vs
\noindent
\underline{\textit{Case 3: $q.l \leq c_{i} \leq q.r$.}}
In this case, $q$ overlaps $c_{i}$, so all intervals in $L^{l}_{i}$ (and $L^{r}_{i}$) overlap $q$.
As mentioned in Section \ref{sec:uniform:ait}, both $u_{i}^{l}$ and $u_{i}^{r}$ have to be traversed in this case.
The next (actually last) task we have to address is to find the intervals that overlap $[q.l,c_{i})$ and $(c_{i},q.r]$ and are maintained at the sub-trees rooted at $u_{i}^{l}$ and $u_{i}^{r}$.
Thanks to the AIT structure, these intervals are guaranteed to be contained in $AL^{r}$ ($AL^{l}$) of $u_{i}^{l}$ and $AL^{l}$ ($AL^{r}$) of $u_{i}^{r}$.
That is, different from the existing data structures for intervals, \textit{we need to traverse no more nodes}.

Focus on $u_{j} = u_{i}^{l}$ (the left child node of $u_{i}$), and for every $x' \in AL^{r}_{j}$, if $q.l \leq x'.r$, then $x' \cap [q.l,c_{i})$ because $x'.r < c_{i} \leq q.r$.
Next, focus on $u_{k} = u_{i}^{r}$ (the right child node of $u_{i}$), and for every $x' \in AL^{l}_{k}$, if $x'.l \leq q.r$, then $x' \cap (c_{i},q.r]$ because $q.l \leq c_{i} < x'.l$.

\begin{figure}
    \centering
    \includegraphics[width=0.95\linewidth]{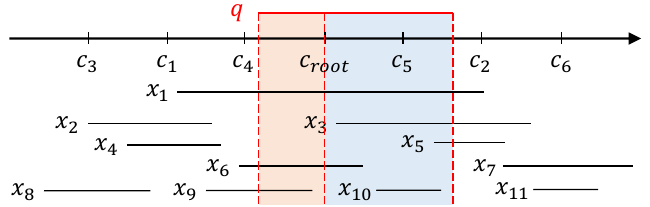}
    \caption{Illustration of case 3: $q.r \leq c_{i} \leq q.r$ in Fig. \ref{fig:ait} ($c_{i} = c_{root}$)}
    \label{fig:algorithm}
\end{figure}

\begin{example}
Fig. \ref{fig:algorithm} illustrates an example of case 3.
It uses the same intervals in Example \ref{example:ait}, so consider the AIT in Fig. \ref{fig:ait}.
The query interval $q$ overlaps $c_{root}$, so it is guaranteed that $q$ overlaps $x_{1}$ and $x_{6}$.
Then, the remaining task is to find the other intervals existing in the shaded space (or the red and blue spaces).
Focus on $u_{1}$ (the left child node of $u_{root}$)
The remaining intervals in the red space are certainly contained in $AL^{r}_{1}$.
Also, focus on $u_{2}$ (the right child node of $u_{root}$).
The remaining intervals in the blue space are also certainly contained in $AL^{l}_{2}$ sequentially.
\end{example}

\subsubsection{\textbf{Algorithm Description}}
Based on the above findings, we design a new algorithm for the IRS problem on interval data, which is summarized in Algorithm \ref{algo:uniform}.
Given a query interval $q$ and a sample size $s$, we access node $u_i$ ($u_{i} = u_{root}$ at initialization) and do the following operations according to the case where $u_{i}$ falls in.

\vs
\noindent
\textit{Case 1.}
As in Example \ref{example:idea}, we focus on $L^{l}_{i}$ and run a binary search with a query point $q.r$ to find the index $j$ such that $L^{l}_{i}[j] \leq q.r < L^{l}_{i}[j+1]$.
From the analysis of case 1 in Section \ref{sec:uniform:algorithn:observation}, each interval $L^{l}_{i}[\textsf{idx}]$ ($1 \leq \textsf{idx} \leq j$) overlaps $q$.
If $j \geq 1$, we make a node record $R_{i}$.
This record maintains (a) an integer, which suggests which list is used (0: $L^{l}$, 1: $L^{r}$, 2: $AL^{l}$, and 3: $AL^{r}$), and (b) a pair of the left and right indices, which represents a sequence of intervals overlapping $q$.
In this case, $R_{i} = \langle 0, (1,j)\rangle$.
Let $\mathcal{R}$ be a set of node records, and $R_{i}$ is added to $\mathcal{R}$.
After that, we traverse $u^{l}_{i}$ (the left child node of $u_{i}$) if it exists.

\vs
\noindent
\textit{Case 2.}
We use $L^{r}_{i}$ and run a binary search with a query point $q.l$ to find the smallest index $j$ such that $q.l \leq L^{r}_{i}[j]$.
This is because, from the analysis of case 2 in Section \ref{sec:uniform:algorithn:observation}, we have $L^{r}_{i}[\textsf{idx}] \cap q$ for $j \leq \textsf{idx} \leq |L^{r}_{i}|$.
If $j \leq |L^{r}_{i}|$, then we make $R_{i} = \langle 1, (j,|L^{r}_{i}|)\rangle$, and this is added to $\mathcal{R}$.
If $u^{r}_{i}$ (the right child node of $u_{i}$) exists, we next traverse it.

\vs
\noindent
\textit{Case 3.}
In this case, all intervals in $L^{l}_{i}$ (and $L^{r}_{i}$) overlap $q$, so we make $R_{i} = \langle 0, (1,|L^{l}_{i}|)\rangle$ and add this to $\mathcal{R}$.
Then, we traverse both $u^{l}_{i}$ and $u^{r}_{i}$ if they exist.
As for $u_{k} = u^{l}_{i}$, we use $AL^{r}_{k}$.
We run a binary search on $AL^{r}_{k}$ with a query point $q.l$ to find the smallest index $j$ such that $q.l \leq AL^{r}_{k}[j]$ as with case 2.
We make $R_{k} = \langle 2, (j,|AL^{r}_{k}|)\rangle$, and this is added to $\mathcal{R}$.
As for $u_{k'} = u^{r}_{i}$, we use $AL^{l}_{k'}$.
We run a binary search on $AL^{l}_{k'}$ with a query point $q.r$ to find the index $j$ such that $AL^{l}_{k'}[j] \leq q.r < AL^{l}_{k'}[j+1]$, as in case 1.
Also, $R_{k'} = \langle 3, (1,j)\rangle$ is added to $\mathcal{R}$.
No more node traversals are required in case 3.

\vs
\noindent
Note that we do not use $AL^{l}$ and $AL^{r}$ in cases 1 and 2 to guarantee correctness because, in these cases, even if $x'.l \leq q.r$ for $x' \in AL^{l}$ (or $q.l \leq x'.r$ for $x' \in AL^{r}$), it is not guaranteed that $x' \cap q$.

Now we obtain $\mathcal{R}$, and recall that each node record in $\mathcal{R}$ conceptually represents a sequence of intervals overlapping $q$.
Let \textsf{idx}$^{l}_{i}$ and \textsf{idx}$^{r}_{i}$ respectively be the left and right indices maintained in $R_{i}$. 
It is important to notice that the sequence size (i.e., $\textsf{idx}^{r}_{i} - \textsf{idx}^{l}_{i} + 1$) is different for each node record.
Therefore, to satisfy equal sampling probability for each interval $x \cap q$, simple random sampling is prohibitive and weighted sampling is required.
We use Walker's alias method to build an alias of $\mathcal{R}$, where the weight of $R_{i}$ is \textsf{idx}$^{r}_{i} -$ \textsf{idx}$^{l}_{i} + 1$.
Then, we randomly pick a weighted sample (i.e., a node record in $\mathcal{R}$) by using the alias.
Assume that this sample is $R_{i}$, and we pick an index from [\textsf{idx}$^{l}_{i},$ \textsf{idx}$^{r}_{i}$] uniformly at random.
From this index and the integer in $R_{i}$, we obtain an interval as a random sample.
This sampling is repeated $s$ times.

\begin{algorithm}[!t]
    \caption{IRS on non-weighted interval data}	\label{algo:uniform}
    \DontPrintSemicolon
    \KwIn {An AIT of $X$, $q$, $s$}
    \KwOut{$S$}
    $S \leftarrow \varnothing$, $\mathcal{R} \leftarrow \varnothing$, $u_{i} \leftarrow u_{root}$\; \label{algo:uniform:begin}
    \While {$1$}
    {
        \uIf{$q.r < c_{i}$}
        {
            $j \gets$ \textsc{Binary-Search}$(L^{l}_{i},q.r)$  \Comment*[r]{\scriptsize \textsf{idx} search}
            \textbf{if} $j \geq 1$ \textbf{then} $\mathcal{R} \gets \mathcal{R} \cup \langle 0, (1,j)\rangle$\;
            \textbf{if} $u^{l}_{i}$ exists \textbf{then} $u_{i} \gets u^{l}_{i}$ \textbf{else break} 
        }
        \uElseIf {$c_{i} < q.l$}
        {
            $j \gets$ \textsc{Binary-Search}$(L^{r}_{i},q.l)$\;
            \textbf{if} $j \leq |L^{r}_{i}|$ \textbf{then} $\mathcal{R} \gets \mathcal{R} \cup \langle 1, (j,|L^{r}_{i}|)\rangle$\;
            \textbf{if} $u^{r}_{i}$ exists \textbf{then} $u_{i} \gets u^{r}_{i}$ \textbf{else break} 
        }
        \Else
        {
            $\mathcal{R} \gets \mathcal{R} \cup \langle 0, (1,|L^{l}_{i}|)\rangle$\;
            \If {$u^{l}_{i}$ exists}
            {
                $u_{k} = u^{l}_{i}$     \Comment*[r]{\scriptsize go to the left child of $u_{i}$}
                 $j \gets$ \textsc{Binary-Search}$(AL^{r}_{k},q.l)$\;
                \textbf{if} $j \leq |AL^{r}_{k}|$ \textbf{then} $\mathcal{R} \gets \mathcal{R} \cup \langle 2, (j,|AL^{r}_{k}|)\rangle$\;
            }
            \If {$u^{r}_{i}$ exists}
            {
                $u_{k} = u^{r}_{i}$    \Comment*[r]{\scriptsize go to the right child of $u_{i}$}
                $j \gets$ \textsc{Binary-Search}$(AL^{l}_{k},q.r)$\;
                \textbf{if} $j \geq 1$ \textbf{then} $\mathcal{R} \gets \mathcal{R} \cup \langle 3, (1,j)\rangle$
            }
            \textbf{break}
        }
    }
    \textbf{if}  $\mathcal{R} = \varnothing$ \textbf{then} return $\varnothing$\;
    $\mathcal{A} \gets$ \textsc{Build-Alias}$(\mathcal{R})$    \Comment*[r]{\scriptsize Building an alias by Walker's method}  \label{algo:uniform:alias}
    \While {$|S| < s$}
    {
        $R_{i} = \langle j, ($\textsf{idx}$^{l}_{i}$, \textsf{idx}$^{r}_{i})\rangle \gets$ \textsc{Weighted-Sampling}$(\mathcal{A})$\;
        \textsf{idx} $\gets$ a random integer in $[\textsf{idx}^{l}_{i}, \textsf{idx}^{r}_{i}]$\;  \label{algo:uniform:sampling}
        \textbf{if} $j = 0$ \textbf{then} $S \gets S \cup \{L^{r}_{i}[\textsf{idx}]\}$\;
        \textbf{else if} $j = 1$ \textbf{then} $S \gets S \cup \{L^{l}_{i}[\textsf{idx}]\}$\;
        \textbf{else if} $j = 2$ \textbf{then} $S \gets S \cup \{AL^{r}_{i}[\textsf{idx}]\}$\;
        \textbf{else} $S \gets S \cup \{AL^{l}_{i}[\textsf{idx}]\}$
    }
    \textbf{return} $S$
\end{algorithm}

\vspace{2.0mm}
\subsubsection{\textbf{Analysis}}
We clarify that Algorithm \ref{algo:uniform} needs $\tilde{O}(s)$ time\footnote{One may have an idea that range trees (by using the 2-dimensional mapping of intervals) can have a similar result.
However, range trees generally assume no duplicated coordinates (or the heights become $O(n)$) and cannot early terminate tree traversal even when case 3 is met.
The AIT structure does not have these issues.}.

\begin{theorem}[\textsc{Time complexity of Algorithm \ref{algo:uniform}}]   \label{theorem:time:uniform}
The time complexity of Algorithm \ref{algo:uniform} is $O(\log^{2}n + s)$.
\end{theorem}

\noindent
\textsc{Proof.}
Assume that node $u_{i}$ has case 1 or 2.
We need $O(\log n)$ time to obtain the corresponding index in $L^{l}_{i}$ or $L^{r}_{i}$, due to the binary search.
Recall that the height of the AIT is $O(\log n)$, and we have cases 1 and 2 at most $O(\log n)$ time, so the total time cost of cases 1 and 2 is $O(\log n) \times O(\log n) = O(\log^{2}n)$.
Next assume that $u_{i}$ has case 3, and we have this case at most once.
The main time cost incurred by case 3 is for the two binary searches on its left and right child nodes, which need $O(\log n)$ time.
Hence, computing $\mathcal{R}$ needs $O(\log^{2}n)$ time in the worst case.

Now we analyze the time cost for sampling an interval $x$ from $\mathcal{R}$.
As seen from the above analysis, we access at most $O(\log n)$ nodes, so $|\mathcal{R}| = O(\log n)$.
That is, we need $O(\log n)$ time to build the alias.
Section \ref{sec:preliminary:sampling} clarifies that sampling a node record from the alias needs $O(1)$ time, and sampling an interval from the node record also needs $O(1)$ time.
Given these results, Algorithm \ref{algo:uniform} needs $O(\log^{2}n) + O(\log n) + O(s) = O(\log^{2}n + s)$ time.  \wsq

\vspace{2.0mm}
We next show that each $x \in q \,\cap\, X$ has equal and independent sampling probability.

\begin{theorem}[\textsc{Correctness of Algorithm \ref{algo:uniform}}]  \label{theorem:correctness}
Algorithm \ref{algo:uniform} guarantees that, in each sampling iteration, any $x \in q \cap X$ is sampled with probability $\frac{1}{|q \cap X|}$.
\end{theorem}

\noindent
\textsc{Proof.}
We first prove that Algorithm \ref{algo:uniform} does not miss any intervals $\in q \cap X$ (i.e., no false negatives) and not have any intervals $\notin q \cap X$ (i.e., no false positives).
Assume that we have only cases 1 and 2.
The pruned sub-trees clearly do not have intervals overlapping $q$.
In addition, for each traversed node $u_{i}$, Algorithm \ref{algo:uniform} evaluates $L^{l}_{i}$ or $L^{r}_{i}$, meaning that all of not-pruned intervals are considered.
Let $\textsf{idx}^{l}_{i}$ and $\textsf{idx}^{r}_{i}$ respectively be the obtained left and right indices of $L^{l}_{i}$ or $L^{r}_{i}$.
The observations in Section \ref{sec:uniform:algorithn:observation} derive the following three facts.
(i) $L^{l}_{i}[j]$ or $L^{r}_{i}[j]$ overlaps $q$ for $\textsf{idx}^{l}_{i} \leq j \leq \textsf{idx}^{r}_{i}$.
(ii) For case 1 and \textsf{idx} such that \textsf{idx} $> \textsf{idx}^{r}_{i}$, $L^{l}_{i}[\textsf{idx}]$ does not overlap $q$.
(iii) Consider case 2 and \textsf{idx} such that \textsf{idx} $< \textsf{idx}^{l}_{i}$, $L^{r}_{i}[\textsf{idx}]$ does not overlap $q$.
Therefore, no false positives and negatives are incurred in cases 1 and 2.

We next remove the above assumption: $u_{i}$ meets case 3.
As all intervals in $L^{l}_{i}$ (and $L^{r}_{i}$) overlap $q$, Algorithm \ref{algo:uniform} does not miss them.
Then, Algorithm \ref{algo:uniform} accesses its left and right child nodes ($u_k$ and $u_{k'}$, respectively).
At each of these nodes, we consider all intervals existing in the sub-tree rooted at it, since we use $AL^{l}_{k'}$ or $AL^{r}_{k}$.
Then, Algorithm \ref{algo:uniform} obtains the left and right indices of $AL^{l}_{k'}$ and $AL^{r}_{k}$.
These indices also satisfy no false positives and negatives, as with cases 1 and 2.
Consequently, we see that the set $\mathcal{R}$ of node records maintains the space where only $q \cap X$ exists.

Let $X(R_{i})$ be a set of intervals that can be sampled by using $R_{i}$, and we have $X(R_{i}) \cap X(R_{j}) = \varnothing$ for any $R_{i}, R_{j} \in \mathcal{R}$.
This is because the AIT structure has (a) $L_{i} \cap L_{j} = \varnothing$ ($L_{i}$ can be $L^{l}_{i}$ and $L^{r}_{i}$) for any two nodes ($u_{i}$ and $u_{j}$) and (b) $L_{i} \cap AL_{j} = \varnothing$ ($AL_{j}$ can be $AL^{l}_{j}$ and $AL^{r}_{j}$) for an arbitrary node $u_{j}$ and every of its ancestor nodes $u_{i}$.
Therefore, the sampling probability of a node record $R_{i}$ is $\frac{\textsf{idx}^{r}_{i} - \textsf{idx}^{l}_{i} + 1}{\sum_{\mathcal{R}}(\textsf{idx}^{r}_{j} - \textsf{idx}^{l}_{j} + 1)} = \frac{\textsf{idx}^{r}_{i} - \textsf{idx}^{l}_{i} + 1}{|q \cap X|}$.
The sampling probability of an interval by using $R_{i}$ is $1/(\textsf{idx}^{r}_{i} - \textsf{idx}^{l}_{i} + 1)$.
Consequently, the sampling probability of an interval $x$ from $\mathcal{R}$ (or $q \cap X$) is $1/|q \cap X|$ for any $x \in q \cap X$.
\wsq

\vs
The above proof clarifies that computing $\mathcal{R}$ provides $|q \cap X|$, so the following is true from Theorem \ref{theorem:time:uniform} and proves a faster range counting time than $O(\sqrt{n})$ time of $k$d-tree, for a sufficiently large $n$.

\begin{corollary}[\textsc{Time complexity of range counting on AIT}]
The time complexity of range counting on an AIT is $O(\log^{2}n)$.
\end{corollary}

\subsection{Reducing the Space Complexity}  \label{sec:uniform:aitv}
Since the space of an AIT is near linear to $n$, reducing it to linear to $n$ is desirable.
We hence consider how to theoretically guarantee $O(n)$ space.
By allowing expected time, we can achieve $O(n)$ space.

\vs
\noindent
\textbf{Pre-processing.}
We partition $X$ into disjoint subsets so that each subset has $\Theta(\log n)$ intervals.
Therefore, we have $\Theta(\frac{n}{\log n})$ subsets, i.e., $X = \bigcup_{\Theta(\frac{n}{\log n})} X_{i}$ and, for arbitrary two subsets $X_{i}$ and $X_{j}$ ($i \neq j$), $X_{i} \cap X_{j} = \varnothing$.
(Any partitioning methods can obtain the theoretical result in this section, so we later show our partitioning method.)
For each subset, we make a \textit{virtual} interval.

\begin{definition}[\textsc{Virtual interval}]   \label{definition:virtual}
Given a subset $X_{i}$ of $X$, the virtual interval of $X_{i}$, $v_{i}$, is defined as: $v_{i}.l = \min_{x \in X_{i}} x.l$ and $v_{i}.r = \max_{x \in X_{i}} x.r$.
\end{definition}

We build an AIT on a set $V$ of virtual intervals, and, since $|V| = \Theta(\frac{n}{\log n})$, we have the following corollary from Theorem \ref{theorem:ait:space}.

\begin{corollary}
Given a set $V$ of virtual intervals such that $|V| = \Theta(\frac{n}{\log n})$, the space complexity of the AIT on $V$ is $O(n)$.    
\end{corollary}

\vs
\noindent
\textbf{Query processing.}
Note that the AIT on $V$ is denoted by \textit{AIT-V}.
To obtain $s$ random intervals by using the AIT-V, we add the following operations to Algorithm \ref{algo:uniform}.
After we sample a virtual interval $v_{j}$ from $R_{i}$, we sample an interval from $X_{i}$ uniformly at random.
(Assume that each $X_{i}$ has an equal number of intervals, which is easy to hold by adding pseudo-intervals to $X_{\frac{n}{\log n}}$ if necessary.)
Let this sample be $x$, and it is not guaranteed that $x \cap q$.
We hence add $x$ into $S$ iff $x \cap q$.

\begin{corollary}[\textsc{Time complexity of our IRS algorithm on AIT-V}]   \label{corollary:aitv}
Our algorithm on an AIT-V needs $O(\log^{2} + s)$ expected time.    
\end{corollary}

\noindent
\textsc{Proof.}
As with Algorithm \ref{algo:uniform}, it needs $O(\log^{2}n)$ time to obtain $\mathcal{R}$.
Definition \ref{definition:virtual} guarantees that if $x \in X_{i}$ has $x \cap q$, $v_{i} \cap q$.
To sample an interval $x$ such that $x \cap q$, we need only a constant number of iterations in expectation \cite{simpler2023aoyama}.
Therefore, we need $O(s)$ expected time to obtain $S$ after obtaining $\mathcal{R}$.    \wsq

\vspace{2.0mm}
It is important to notice that, in AIT-V, each interval $x$ which can be sampled from $\mathcal{R}$ has equal sampling probability.  
This is trivial from the proof of Theorem \ref{theorem:correctness}.

\vs
\noindent
\textbf{How to partition $X$.}
Any disjoint partitioning and Definition \ref{definition:virtual} provide Corollary \ref{corollary:aitv}, but minimizing the number of failure iterations (i.e., sampling an interval $\notin q \cap X$) is crucial for reducing practical time.
A reasonable idea is to minimize the difference between $v_{i}$ and each $x_{j} \in X_{i}$.
Then, we can formulate an optimization problem which minimizes $\sum_{v_{i} \in V}\sum_{x_{j} \in X_{i}}(|v_{i}.l - x_{j}.l| + |v_{i}.r - x_{j}.r|)$ such that $|X_{k}| = \log n$ $(k \in [1, \frac{n}{\log n}])$.
This is equivalent to the $p$-median problem, one of the well-known facility location problems, with capacity constraint, which is NP-hard \cite{daskin2015p}.
(In our case, $v_{i} \in \mathbb{R}^{2}$ for each $i \in [1, \frac{n}{\log n}]$.)

\begin{figure}[!t]
    \begin{center}
        \subfigure[Book]{%
            \includegraphics[width=0.45\linewidth]{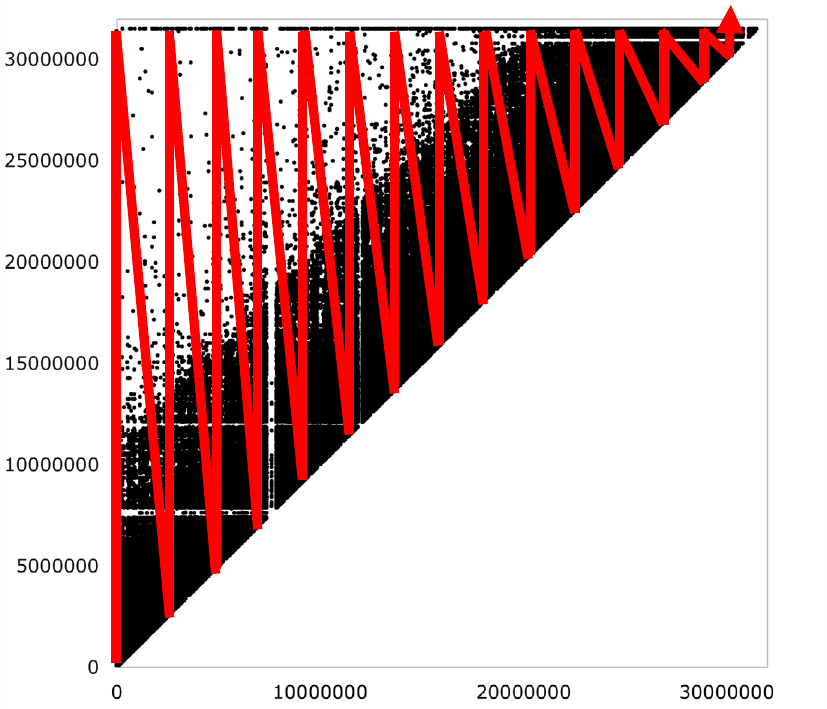}    \label{fig:distribution:book}}
        \subfigure[BTC]{%
    	\includegraphics[width=0.45\linewidth]{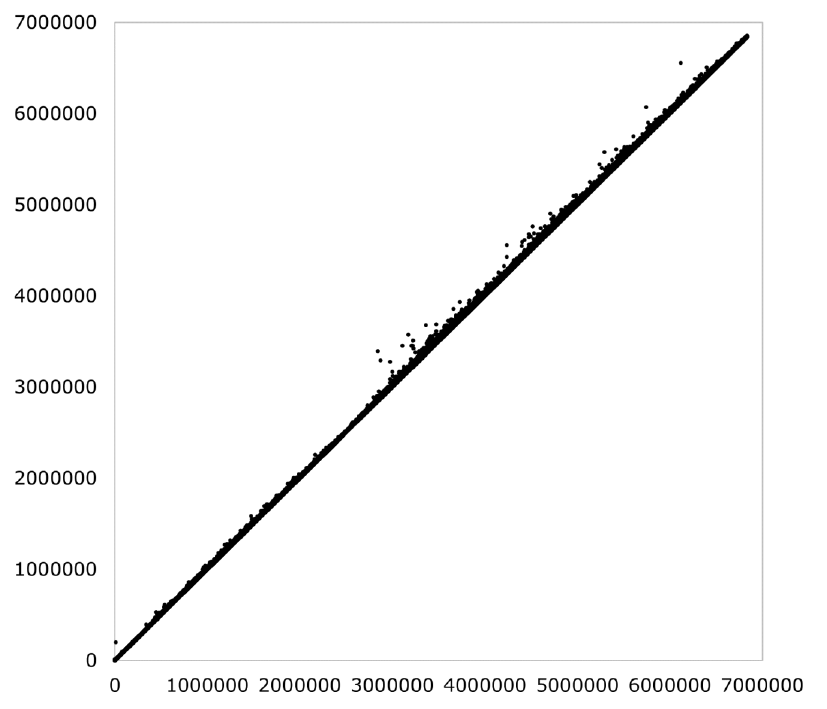}     \label{fig:distribution:btc}}
        \caption{Distribution of intervals and rough z-curve (for Book dataset)}
        \label{fig:distribution}
    \end{center}
\end{figure}

We do not solve the above optimization problem but use its concept: the intervals in $X_{i}$ should be similar to each other.
To this end, we employ a pair sort, which sorts $X$ in ascending order of left-endpoint and breaks ties by sorting them in ascending order of right-endpoint.
We partition $X$ based on this sort order.
Fig. \ref{fig:distribution} illustrates the distributions of the intervals in the real-world datasets Book and BTC, where each $x = [x.l, x.r]$ is mapped into two-dimensional coordinates $(x.l, x.r)$ (a query interval $q$ is mapped to the rectangle $[-\infty, q.r] \times [q.l, \infty]$).
In Fig. \ref{fig:distribution:book}, the red line (roughly) represents the sort order.
This figure shows that pair sorting is equivalent to making a z-curve, which maps multi-dimensional points to one-dimensional values while preserving locality.

This approach has two merits: it (i) needs only $O(n\log n)$ additional cost in pre-processing and (ii) functions well in practice for both datasets with large-length intervals (e.g., Book) and with short-length intervals (e.g., BTC).
The only possible concern is the partitions with the curves in the former case, because they lose the locality compared with those with no curves.
However, each partition contains only $\Theta(\log n)$ intervals, so the impact of the curves is negligible.
Our experiments observe that the number of sampling iterations in AIT-V is almost $s$.
For example, when $s = 1000$, it is 1087 (1020) on average in Book (BTC).

\subsection{Updates}
We show how to update the AIT structure when an insertion or a deletion is given.
(Note that we do not assume frequent updates, and dynamic intervals with high-frequency updates are not the scope of this work.)

\vs
\noindent
\textbf{Insertion.}
Given a new interval, we traverse the AIT nodes in the same way as Algorithm \ref{algo:uniform}, i.e., we traverse the nodes until we have case 3.
When a traversed node $u_{i}$ has case 1 (or 2), we update $AL^{l}_{i}$ and $AL^{r}_{i}$ and go to its left (or right) child node.
If $u_{i}$ does not have a left (or right) child node, we create it, and the insertion is over.
(If the height exceeds $\log n$, we rebuild the AIT to hold the theoretical performance guarantee.)
When $u_{i}$ has case 3, we update $L^{l}_{i}$ and $L^{r}_{i}$, and then the insertion is over.

Although the above approach correctly updates the AIT structure, it needs many sorting operations for a single insertion.
To alleviate this issue, we use batch insertion, where the batch size is $O(\log^{2}n)$.
Given an insertion, we add it to an insertion pool and update the AIT only when the pool contains $O(\log^{2}n)$ intervals.
In this case, we update the AIT similarly to the above one-by-one approach.
A difference is that we first insert all new intervals in the pool to the corresponding lists (but sorting is not done at this time) while marking the corresponding nodes, and then we sort the lists in the marked nodes.
This approach can reduce the amortized insertion cost while keeping the theoretical performance of Algorithm \ref{algo:uniform}.
(When the pool has some new insertions but the AIT is not updated, Algorithm \ref{algo:uniform} additionally accesses the intervals in the pool, which needs $O(\log^{2}n)$ time.)

\vs
\noindent
\textbf{Deletion.}
When an interval is removed, we delete it from the corresponding lists of the AIT.
The node traversal method is the same as that for the one-by-one insertion.
When a node maintains no intervals, we remove it from the AIT.
As this deletion does not incur sorting, we do not consider batch updates.

\section{Weighted Interval Case} \label{sec:weighted}
This section solves Problem \ref{problem:weighted} and shows that there exists an exact algorithm that needs $\tilde{O}(s)$ time and $\tilde{O}(n)$ space.

\vs
\noindent
\textbf{Main idea.}
Our basic idea is to extend the AIT structure and Algorithm \ref{algo:uniform} for weighted intervals.
An intuitive method is to make an alias while considering the weights of $X(R_{i})$ for $R_{i} \in \mathcal{R}$ (recall that $X(R_{i})$ is a set of intervals that can be sampled by using $R_{i}$).
However, to obtain the weights, we need to access all intervals in $X(R_{i})$.
Since $|X(R_{i})|$ can be as large as $O(n)$, this approach incurs $O(n)$ time.
Our idea for avoiding this issue is to associate cumulative sums of weights with the sorted lists of each node.
This can be prepared in the pre-processing phase.
Therefore, for a given query, we need little computational cost to obtain the total weight of the sequence of the intervals defined by $\textsf{idx}^{l}$ and $\textsf{idx}^{r}$.
We exploit this idea to achieve an $\tilde{O}(s)$ time algorithm.

\subsection{Augmented Weighted Interval Tree}
We extend the AIT structure to efficiently deal with weighted intervals, based on the above idea.
This extended AIT is called AWIT (Augmented Weighted Interval Tree).
In addition to the components in Section \ref{sec:uniform:ait:structure}, each node $u_{i}$ of an AWIT has the following four arrays:
\begin{itemize}
    \setlength{\leftskip}{-2.5mm}
    \item   $W^{l}_{i}$: an array that maintains the cumulative sums of weights of the intervals in $L^{l}_{i}$.
            Its $j$-th element, $W^{l}_{i}[j]$, is $\sum_{k = 1}^{j}w(L^{l}_{i}[k])$. 
    \item   $W^{r}_{i}$: an array whose $j$-th element, $W^{r}_{i}[j]$, is $\sum_{k = 1}^{j}w(L^{r}_{i}[k])$.
    \item   $AW^{l}_{i}$: an array whose $j$-th element, $AW^{l}_{i}[j]$, is $\sum_{k = 1}^{j}w(AL^{l}_{i}[k])$. 
    \item   $AW^{r}_{i}$: an array whose $j$-th element, $AW^{r}_{i}[j]$, is $\sum_{k = 1}^{j}w(AL^{r}_{i}[k])$.
\end{itemize}

The above definition suggests that each node of the AWIT needs approximately double-space of that of the AIT, so Theorem \ref{theorem:ait:space} derives the following corollary:

\begin{corollary}[\textsc{Space complexity of AWIT}]    \label{corollary:awit:space}
The space complexity of an AWIT is $O(n\log n)$.
\end{corollary}

The main merit of the above arrays is that only $O(1)$ time is needed to obtain the total weight of intervals which can be sampled from $R_{i}$.
Hence, we can still build an alias of $\mathcal{R}$ in $O(\log n)$ time.
Furthermore, when we sample a random weighted interval from $R_{i}$, we do not need to build any data structures.

\vs
\noindent
\textbf{Discussion.}
Since the above four arrays are based on the sorted intervals in the corresponding lists, even a single insertion/deletion may incur significant updates in the four arrays.
We therefore do not consider updates in Section \ref{sec:weighted}, and designing a data structure that supports efficient IRS query processing and updates on weighted interval data is a future work.

In Problem \ref{problem:weighted}, the sampling probability of a weighted interval is defined based on $q \cap X$.
This suggests that no false positives are allowed, so the idea of AIT-V (see Section \ref{sec:uniform:aitv}) is not available.

\subsection{IRS Algorithm for Weighted Intervals}   \label{sec:weighted:algorithm}
Our IRS algorithm for weighted intervals follows Algorithm \ref{algo:uniform}.
The differences are summarized below:
\begin{itemize}
    \setlength{\leftskip}{-2.5mm}
    \item   We use an AWIT.
    \item   Consider building an alias (line \ref{algo:uniform:alias}).
            For each $R_{i} \in \mathcal{R}$, we use $W^{l}_{i}[\textsf{idx}^{r}_{i}]$, $W^{r}_{i}[\textsf{idx}^{r}_{i}] - W^{r}_{i}[\textsf{idx}^{l}_{i} - 1]$, $AW^{r}_{i}[\textsf{idx}^{r}_{i}] - AW^{r}_{i}[\textsf{idx}^{l}_{i} - 1]$, or $AW^{l}_{i}[\textsf{idx}^{r}_{i}]$ as the weight of $R_{i}$ when we use $L^{l}_{i}$, $L^{r}_{i}$, $AL^{l}_{i}$, or $AL^{r}_{i}$, respectively.
    \item   When sampling a random \textsf{idx} from $[\textsf{idx}^{l}_{i}, \textsf{idx}^{r}_{i}]$ (line \ref{algo:uniform:sampling}), we use the cumulative sum method (see Section \ref{sec:preliminary:sampling}) with $(\textsf{idx}^{l}_{i}, \textsf{idx}^{r}_{i})$ and the corresponding array.
\end{itemize}

One may consider that the last difference can be replaced with Walker's alias method.
If we do this, we need to build an alias on $R_{i}$, which again needs to access all intervals in $X(R_{i})$.
As described before, this is prohibitive.
On the other hand, the cumulative sum method does not need to build any data structures online, so we can avoid the $O(n)$ time.
From this method and Theorem \ref{theorem:time:uniform}, we have:

\begin{corollary}[\textsc{Time complexity of our algorithm on AWIT}]
The time complexity of our IRS algorithm for weighted intervals is $O(\log^{2}n + s\log n)$ time.
\end{corollary}

\noindent
From Theorem \ref{theorem:correctness} and the cumulative sum method, moreover, it is straightforward to see that, in the above algorithm, the sampling probability of any $x \in q \cap X$ is $\frac{{w(x})}{\sum_{x' \in q \cap X} w(x')}$.

Now we have introduced our theoretical results in this paper and the flexible extensions of the AIT structure.
In the next section, we show that the efficiencies of our algorithms are guaranteed not only theoretically but also practically.

\section{Experiment}    \label{sec:experiment}
This section reports our experimental results.
All experiments were conducted on Ubuntu 22.04 LTS machine with 2.2GHz Intel Core i9-13950HX processor and 128GB RAM.

\subsection{Setting}
\noindent
\textbf{Datasets.}
We used four real-world large datasets, Book \cite{implementation2022hint}, BTC \cite{btc}, Renfe \cite{renfe}, and Taxi \cite{taxi}.
Table \ref{tab:stats} shows the statistics of each dataset.
Book contains the borrowing periods of books in Aarhus libraries.
BTC is a set of historical price intervals of Bitcoin, and we used low and high prices as the left- and right-endpoints, respectively.
Renfe and Taxi are respectively Spanish rail and NYC taxi trip data.
For Renfe (Taxi), we used departure (pick-up) time and arrival (drop) time as the left- and right-endpoints, respectively.
When we considered weighted intervals, we assigned a random weight in $[1, 100]$ to each interval.

\vs
\noindent
\textbf{Queries.}
We generated 1,000 queries for each experiment.
The left-endpoint of a query was chosen from the corresponding domain uniformly at random, and its interval length was 8\% of the domain size by default, which was used to set the right-endpoint.
In addition, $s = 1,000$ by default \cite{xie2021spatial}.

\vs
\noindent
\textbf{Evaluated algorithms.}
We used the following algorithms.
\begin{itemize}
    \setlength{\leftskip}{-3.0mm}
    \item   Interval tree \cite{edelsbrunner1980dynamic}:
            This algorithm first searches for all intervals overlapping $q$ and randomly samples $s$ intervals from them.
    \item   \hint \cite{christodoulou2022hint,christodoulou2023hint,implementation2022hint}:
            This took the same approach as Interval tree.
            (Literatures \cite{christodoulou2022hint,christodoulou2023hint} show that \hint significantly outperforms the other state-of-the-art interval search algorithms \cite{kaufmann2013timeline,behrend2019period}, so we did not consider these algorithms.)
    \item   KDS \cite{xie2021spatial}:
            This is the state-of-the-art spatial independent range sampling algorithm.
            (Literature \cite{xie2021spatial} reports that the other algorithms proposed in \cite{xie2021spatial} are outperformed by KDS.)
    \item   AIT: Our algorithm on an AIT (Section \ref{sec:uniform:algorithm}).
    \item   AIT-V: Our algorithm on an AIT-V (Section \ref{sec:uniform:aitv}).
    \item   AWIT: Our algorithm on an AWIT (Section \ref{sec:weighted:algorithm}).
\end{itemize}
The above algorithms were single-threaded, implemented in C++, and compiled by g++ 11.3.0 with -O3 flag\footnote{The source codes of our algorithms are available at \url{https://github.com/amgt-d1/IRS-interval}}.
In the case where intervals were non-weighted, we evaluated Interval tree, HINT$^{m}$, KDS, AIT, and AIT-V.
In the case where intervals were weighted, we evaluated Interval tree, HINT$^{m}$, KDS, and AWIT.
Actually, KDS cannot solve Problem \ref{problem:weighted} correctly, because it incurs false positives and does not guarantee $\frac{w(x)}{\sum_{x_{i} \in q \cap X} w(x_{i})}$ sampling probability.
We used it to highlight the time difference between AWIT and KDS.

\begin{table}[!t]
    \centering
    \caption{Dataset statistics}
    \label{tab:stats}
    \begin{tabular}{lrrrrr} \toprule
        Dataset     & Book          & BTC       & Renfe         & Taxi          \\ \midrule
        Cardinality & 2,295,260     & 2,538,921 & 38,753,060    & 106,685,540   \\
        Domain size & 31,507,200    & 6,876,400 & 52,163,400    & 79,901,357    \\
        Min length  & 3,600         & 1         & 1,320         & 1             \\
        Med. length & 1,458,000     & 937       & 9,120         & 663           \\
        Max length  & 31,406,400    & 547,077   & 44,700        & 2,618,881     \\ \bottomrule
    \end{tabular}
\end{table}

\subsection{Result on Non-weighted Intervals}
\noindent
\textbf{Pre-processing time and memory usage.}
Table \ref{tab:preprocessing} shows the pre-processing time of each algorithm.
AIT needs longer pre-processing times than the others.
However, it is acceptable since pre-processing is done only once, and less than 5 minutes is needed even for about 100 million intervals.
AIT-V alleviates pre-processing time, since it uses only $\frac{n}{\log n}$ (virtual) intervals.
Table \ref{tab:memory} shows the memory usage of each algorithm, and the result is similar to the pre-processing time.
Although AIT needs about 30GB for Taxi (about 100 million intervals), this is easy to fit into modern main memory.
If this can be a concern, AIT-V is preferable because it consumes less memory (and yields fast query processing time, as shown later).

\begin{table}[!t]
    \centering
    \caption{Pre-processing time [sec] (non-weighted case)}
    \label{tab:preprocessing}
    \begin{tabular}{lrrrr} \toprule
        Dataset         & Book  & BTC   & Renfe     & Taxi      \\ \midrule
        Interval tree   & 1.45  & 2.93  & 52.62     & 147.19    \\ 
        \hint           & 0.60  & 0.20  & 3.26      & 4.67      \\
        KDS             & 2.15  & 3.43  & 36.16     & 210.36    \\
        AIT             & 3.02  & 7.00  & 103.52    & 274.02    \\
        AIT-V           & 0.26  & 0.28  & 3.91      & 9.40      \\ \bottomrule
    \end{tabular}
\end{table}
\begin{table}[!t]
    \centering
    \caption{Memory usage [GB] (non-weighted case)}
    \label{tab:memory}
    \begin{tabular}{lrrrr} \toprule
        Dataset         & Book  & BTC   & Renfe     & Taxi  \\ \midrule
        Interval tree   & 0.17  & 0.22  & 2.26      & 6.27  \\ 
        \hint           & 0.10  & 0.06  & 0.53      & 1.29  \\
        KDS             & 0.29  & 0.32  & 4.84      & 13.34 \\
        AIT             & 0.30  & 0.78  & 8.12      & 29.88 \\
        AIT-V           & 0.03  & 0.05  & 0.66      & 1.73  \\ \bottomrule
    \end{tabular}
\end{table}
\begin{figure}[!t]
    \begin{center}
        \subfigure[AIT (pre-processing time)]{%
            \includegraphics[width=0.47\linewidth]{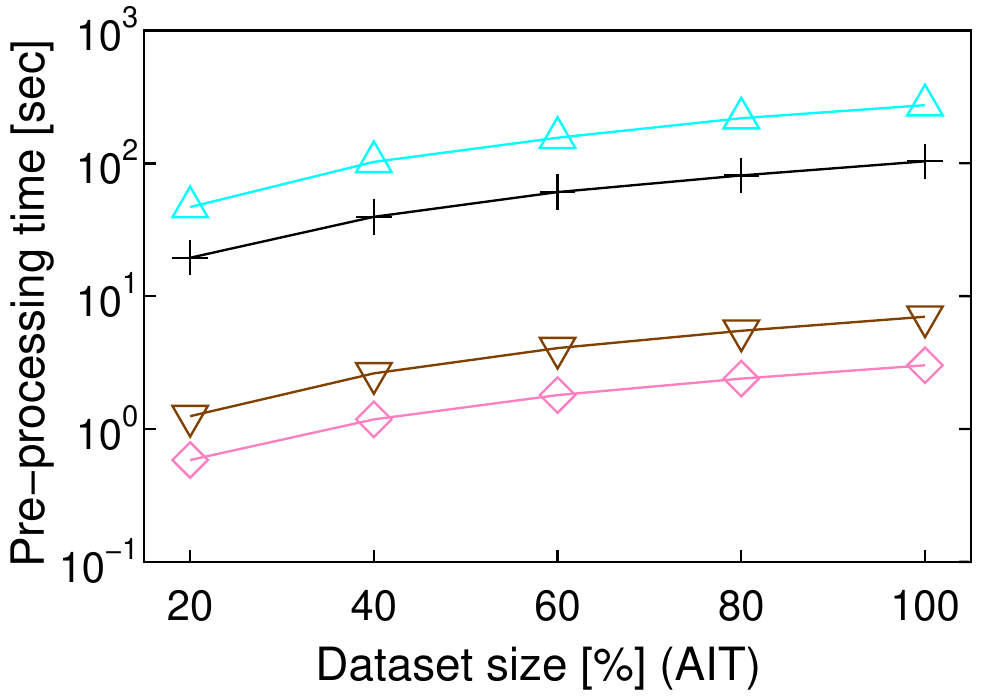}     \label{fig:preprocee-ait}}
        \subfigure[AIT-V (pre-processing time)]{%
    	\includegraphics[width=0.47\linewidth]{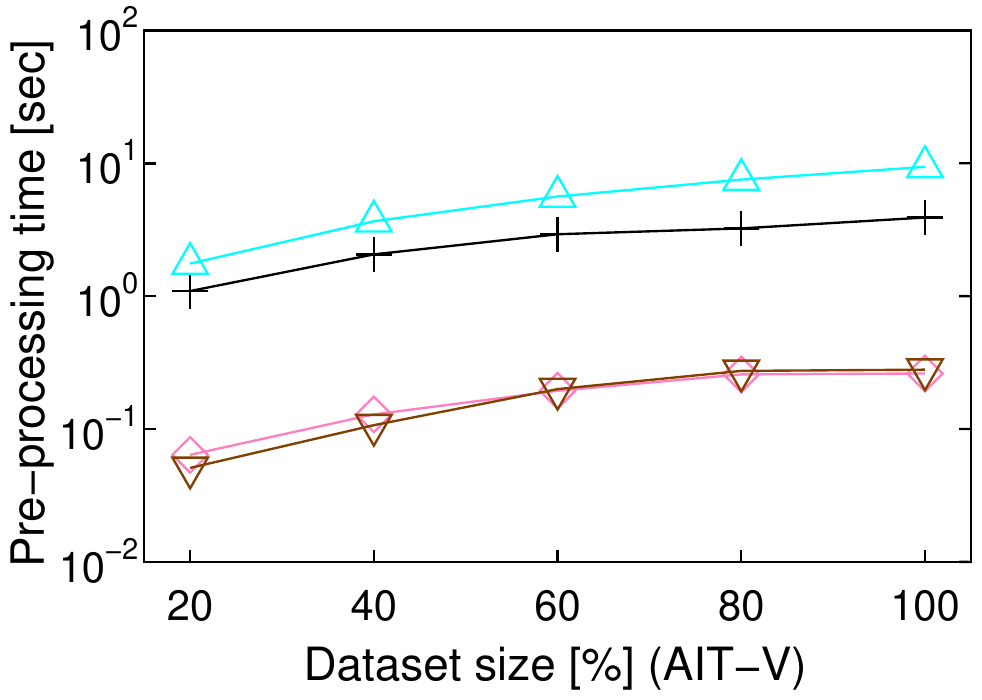}    \label{fig:preprocee-aitv}}
        \subfigure[AIT (memory)]{%
            \includegraphics[width=0.47\linewidth]{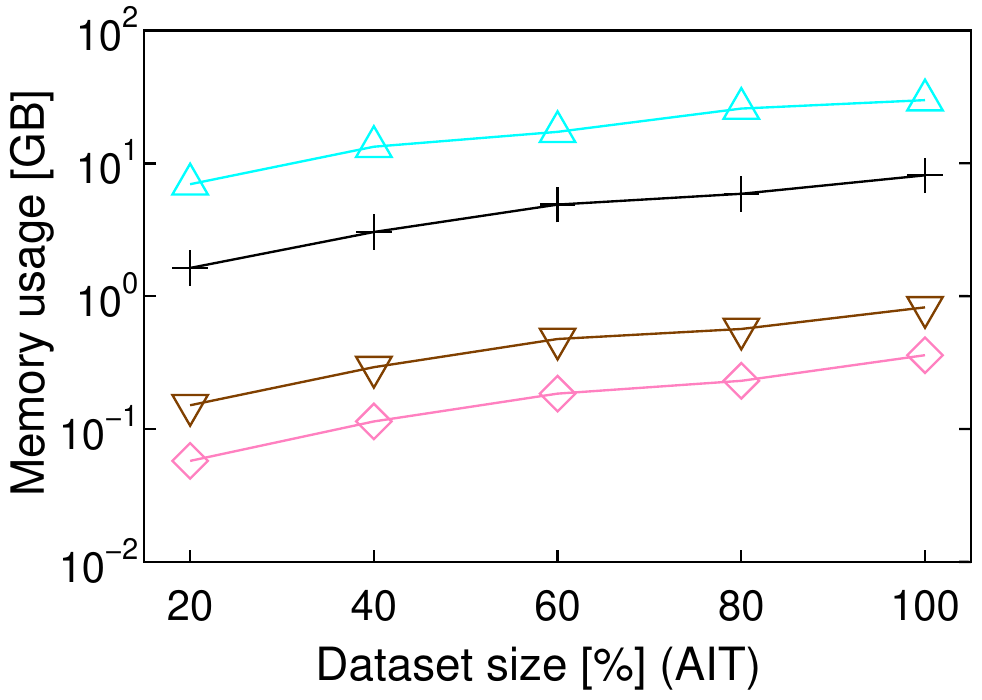}     \label{fig:memory-ait}}
        \subfigure[AIT-V (memory)]{%
    	\includegraphics[width=0.47\linewidth]{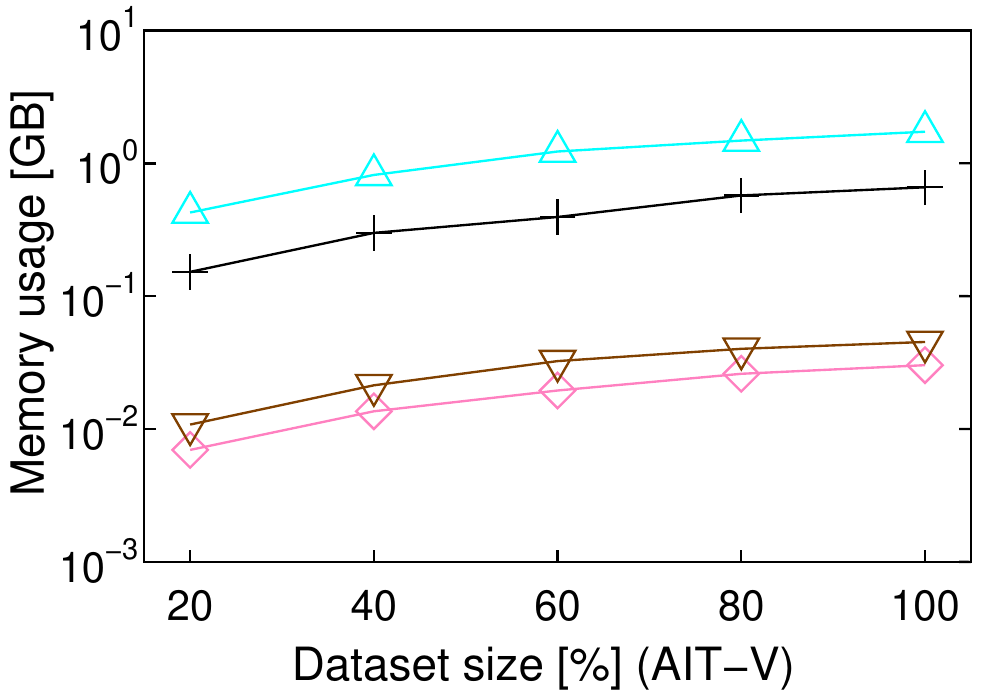}    \label{fig:memory-aitv}}
        \caption{Pre-processing time [sec] and memory usage [GB] of AIT and AIT-V.
        \textcolor{pink}{$\diamond$}, \textcolor{brown}{$\triangledown$}, $+$, and \textcolor{cyan}{$\triangle$} respectively show the result on \textcolor{pink}{Book}, \textcolor{brown}{BTC}, Renfe, and \textcolor{cyan}{Taxi}.}
        \label{fig:preprocess}
    \end{center}
\end{figure}

Fig. \ref{fig:preprocess} depicts how the pre-processing times and memory usages of our algorithms scale to the dataset size.
(We varied the dataset size by random sampling, and plots are log-scale.)
We see that the pre-processing time of each of our algorithms scales linearly to $n$.
Moreover, Fig. \ref{fig:memory-ait} shows that the memory usage of AIT also scales linearly to $n$, which suggests that its practical space performance is better than its theoretical guarantee.
As AIT-V requires $O(n)$ space theoretically, its practical performance follows this result, as shown in Fig. \ref{fig:memory-aitv}.

\vs
\noindent
\textbf{Comparison with existing techniques.}
We next investigate the query processing performance of each algorithm.
Tables \ref{tab:candidate-time} and \ref{tab:sampling-time} respectively show the average time to compute the candidate (i.e., $|q \cap X|$ for Interval tree and \hint, $\mathcal{R}$ for our algorithms, and a superset of $\cup_{\mathcal{R}}X(R_{i})$ for KDS) and the average sampling time.
Interval tree and \hint incur significant time to compute the candidate, since they need to search all intervals in $q \cap X$.
After they obtain $q \cap X$, they run simple random sampling, so their sampling times are the shortest.
KDS alleviates the candidate computation time but requires the longest sampling time, because it needs to use $O(\sqrt{n})$ nodes.

AIT and AIT-V are much better balanced w.r.t. candidate computation and sampling times than the others.
They can obtain $\mathcal{R}$ quite fast, and their sampling times are also short.
For example, on Taxi, AIT is about 3131 (51) times faster than \hint (KDS) w.r.t. total running time (the sum of candidate computation and sampling times).
AIT-V requires a longer sampling time than AIT, because it needs to evaluate whether a sampled interval $x$ has $x \cap q$, whereas AIT does not need this evaluation.
This result suggests the trade-off relationship between time and memory usage.

\begin{table}[!t]
    \centering
    \caption{Candidate computation time [microsec]}
    \label{tab:candidate-time}
     \begin{tabular}{lrrrr} \toprule
        Dataset         & Book      & BTC       & Renfe     & Taxi      \\ \midrule
        Interval tree   & 4353.58   & 3345.17   & 76304.50  & 177287.52 \\ 
        \hint           & 4115.27   & 2183.65   & 34264.49  & 131061.57 \\
        KDS             & 105.29    & 16.37     & 9.40      & 44.24     \\
        AIT             & 0.83      & 0.37      & 1.20      & 2.08      \\
        AIT-V           & 0.02      & 0.01      & 0.94      & 1.01      \\ \bottomrule
    \end{tabular}
\end{table}
\begin{table}[!t]
    \centering
    \caption{Sampling time [microsec] (non-weighted case).
    Alias building time is included.}
    \label{tab:sampling-time}
     \begin{tabular}{lrrrr} \toprule
        Dataset                 & Book      & BTC       & Renfe     & Taxi      \\ \midrule
        Interval tree \& \hint  & 4.79      & 7.39      & 19.81     & 27.43     \\
        KDS                     & 420.13    & 459.70    & 925.84    & 1070.09   \\
        AIT                     & 23.88     & 21.74     & 35.68     & 39.77     \\
        AIT-V                   & 58.14     & 56.00     & 155.93    & 180.95    \\ \bottomrule
    \end{tabular}
\end{table}

\begin{figure}[!t]
    \begin{center}
        \subfigure[Book]{%
    	\includegraphics[width=0.47\linewidth]{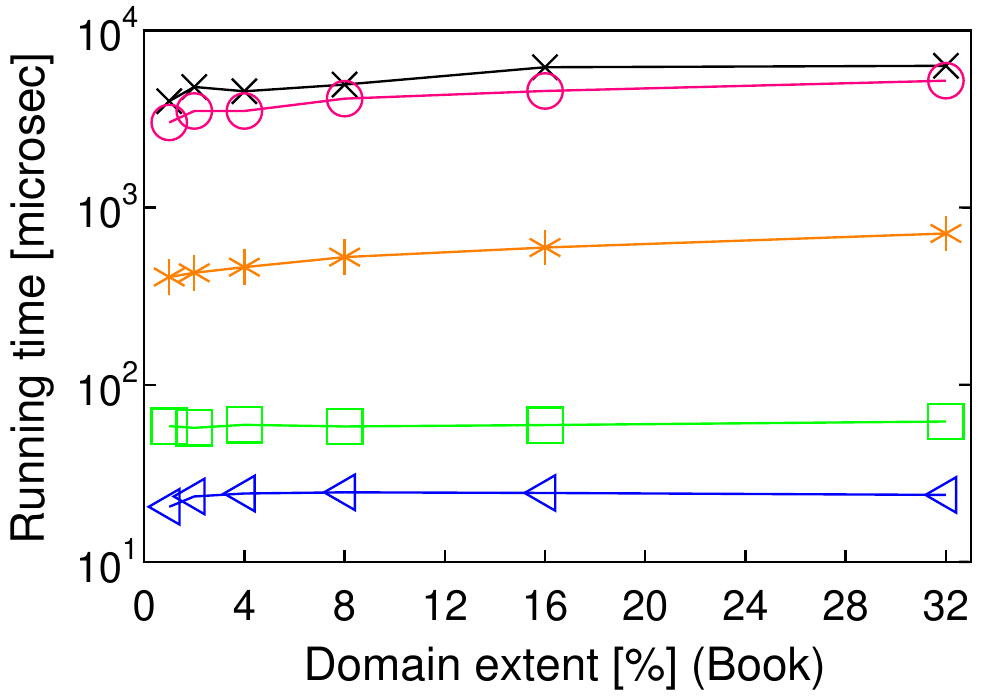}      \label{fig:extent_book}}
        \subfigure[BTC]{%
    	\includegraphics[width=0.47\linewidth]{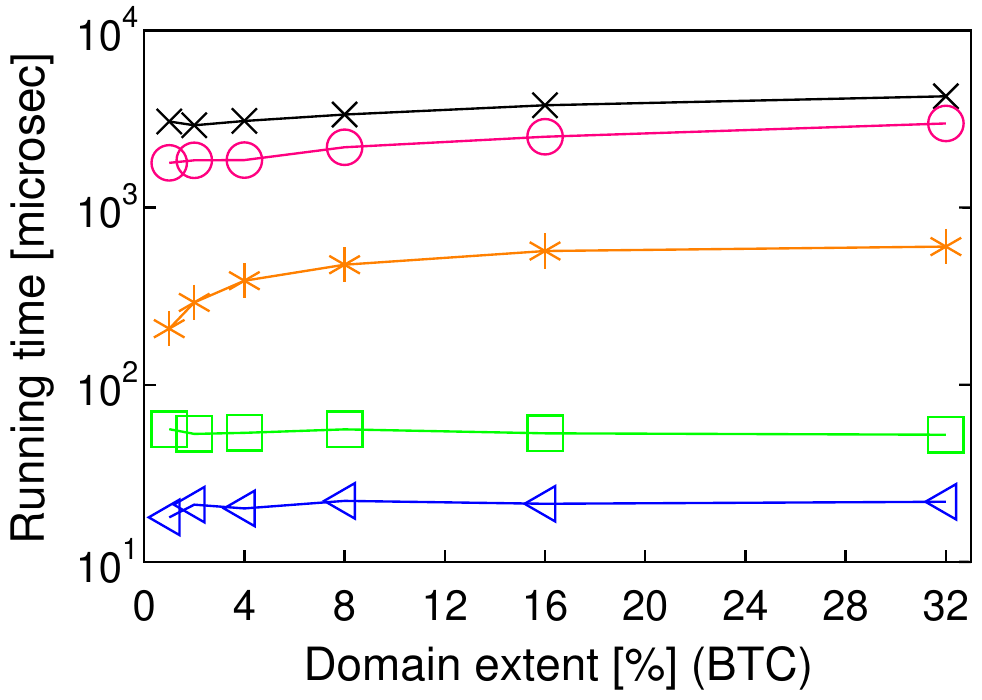}      \label{fig:extent_btc}}
        \subfigure[Renfe]{%
            \includegraphics[width=0.47\linewidth]{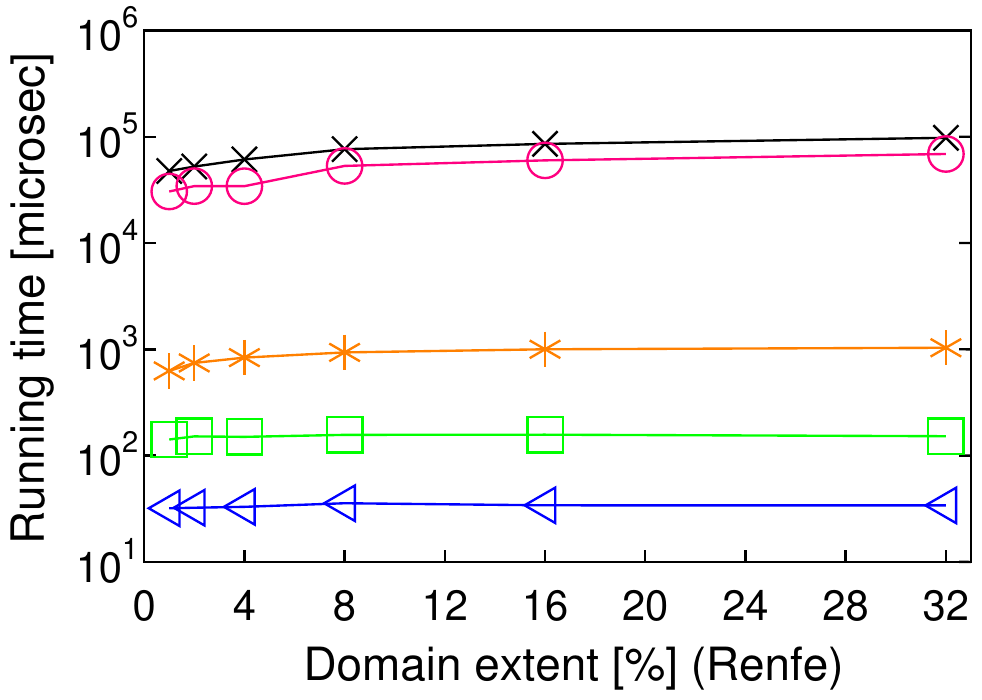}     \label{fig:extent_renfe}}
        \subfigure[Taxi]{%
    	\includegraphics[width=0.47\linewidth]{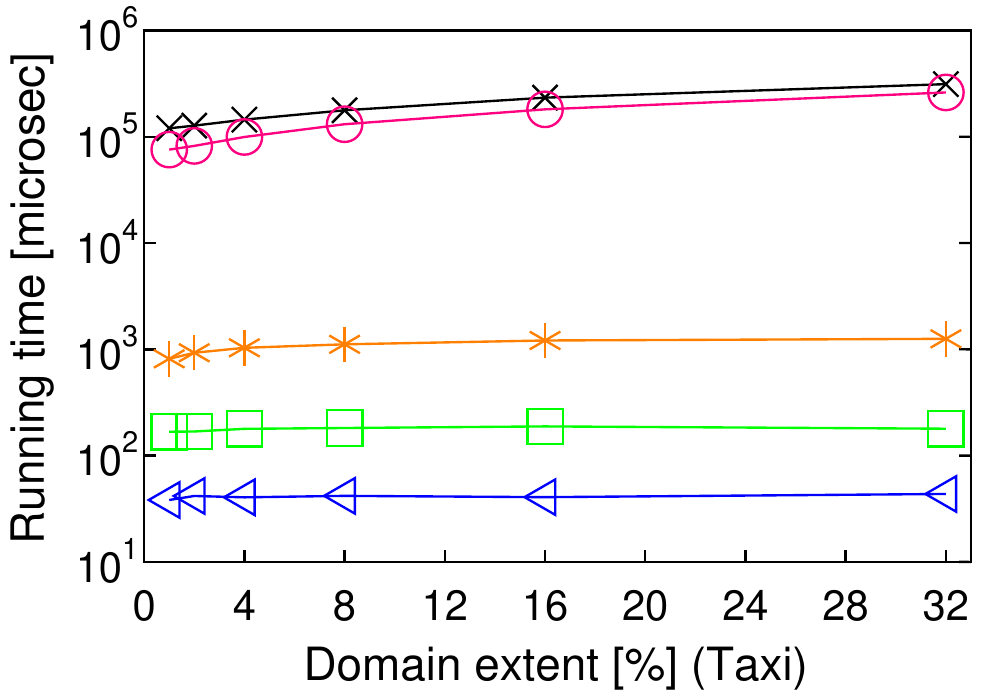}      \label{fig:extent_taxi}}
        \caption{Impact of domain extent (or query interval length) in non-weighted case.
        $\times$ shows Interval tree, \textcolor{magenta}{$\circ$} shows \textcolor{magenta}{\hint}, \textcolor{orange}{$\ast$} shows \textcolor{orange}{KDS}, \textcolor{blue}{$\triangleleft$} shows \textcolor{blue}{AIT}, and \textcolor{green}{$\square$} shows \textcolor{green}{AIT-V}.}
        \label{fig:extent}
    \end{center}
\end{figure}

\vs
\noindent
\textbf{Impact of query interval length.}
We study the impact of range size, i.e., $q.r - q.l$.
Fig. \ref{fig:extent} shows the average running times with different query interval lengths (that are equal to the domain extent).
Interval tree and \hint need longer running time as the domain extent (query interval size) becomes larger.
This is trivial, as larger domain extents provide larger $|q \cap X|$, and they incur $\Omega(|q \cap X|)$ time.
KDS also needs a bit longer time with the increase of query interval length, which is consistent with the result in \cite{xie2021spatial}.
On the other hand, our algorithms are not affected by query interval length, and this result confirms the clear advantage of our algorithms against the competitors.

\begin{figure}[!t]
    \begin{center}
        \subfigure[Book]{%
    	\includegraphics[width=0.47\linewidth]{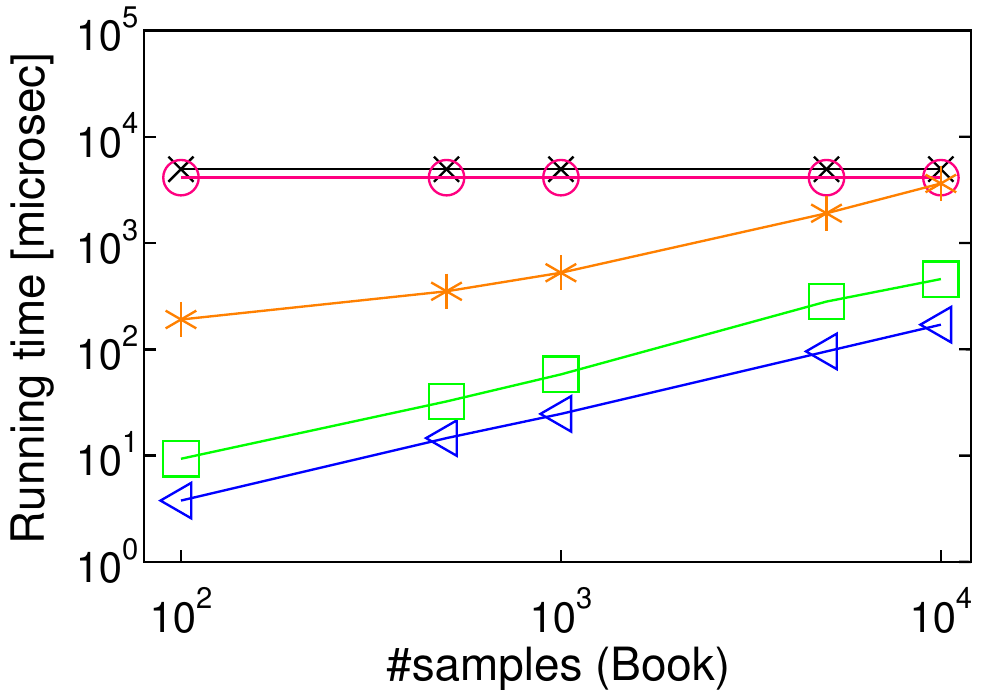}      \label{fig:sample_book}}
        \subfigure[BTC]{%
    	\includegraphics[width=0.47\linewidth]{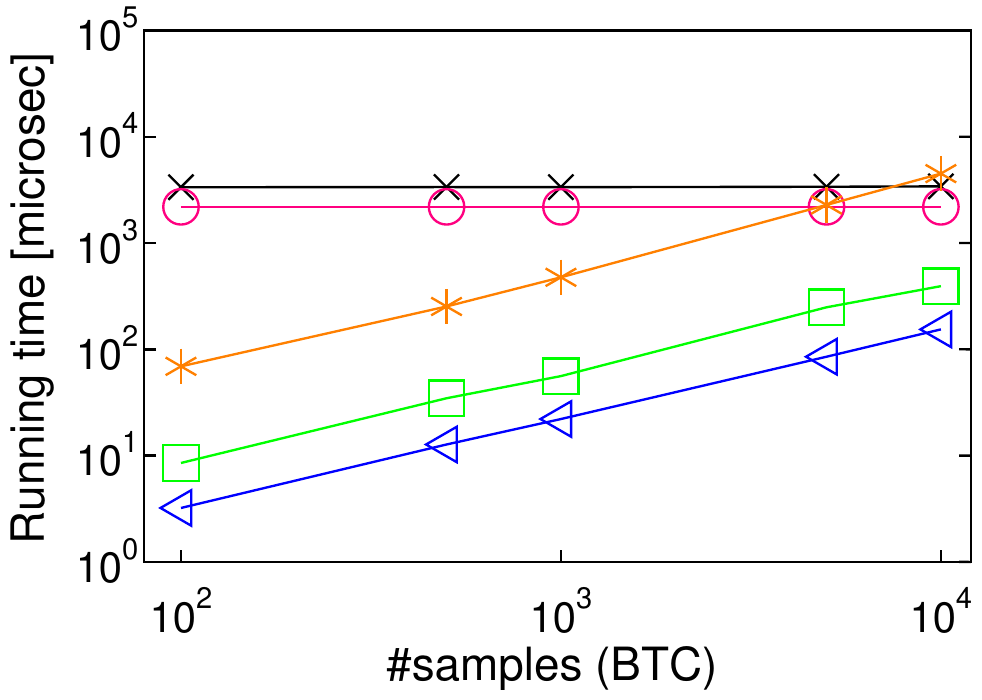}      \label{fig:sample_btc}}
        \subfigure[Renfe]{%
            \includegraphics[width=0.47\linewidth]{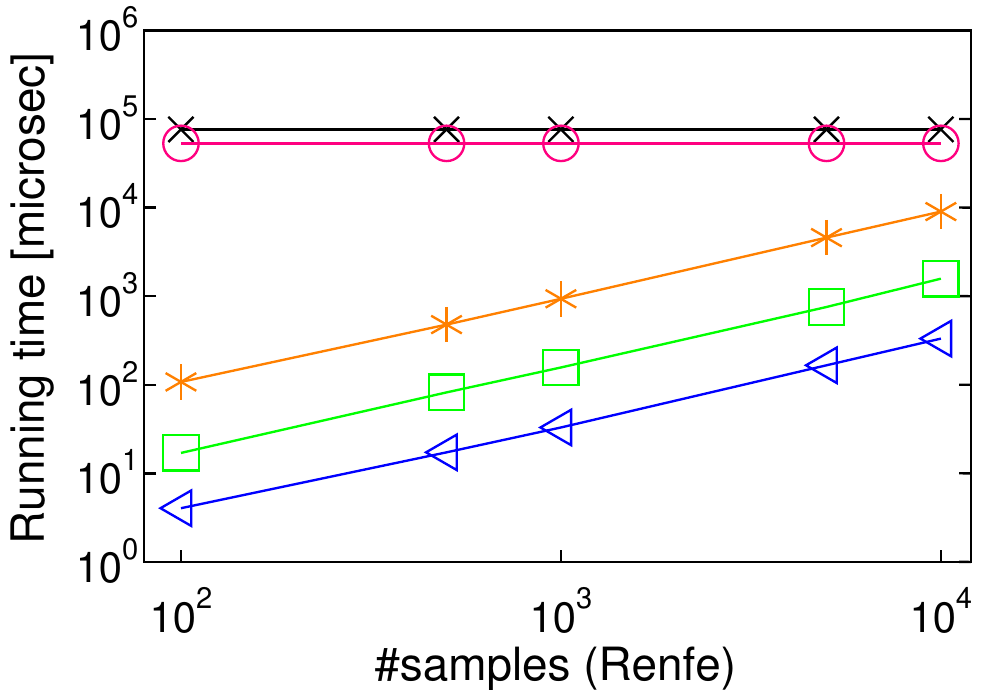}     \label{fig:sample_renfe}}
        \subfigure[Taxi]{%
    	\includegraphics[width=0.47\linewidth]{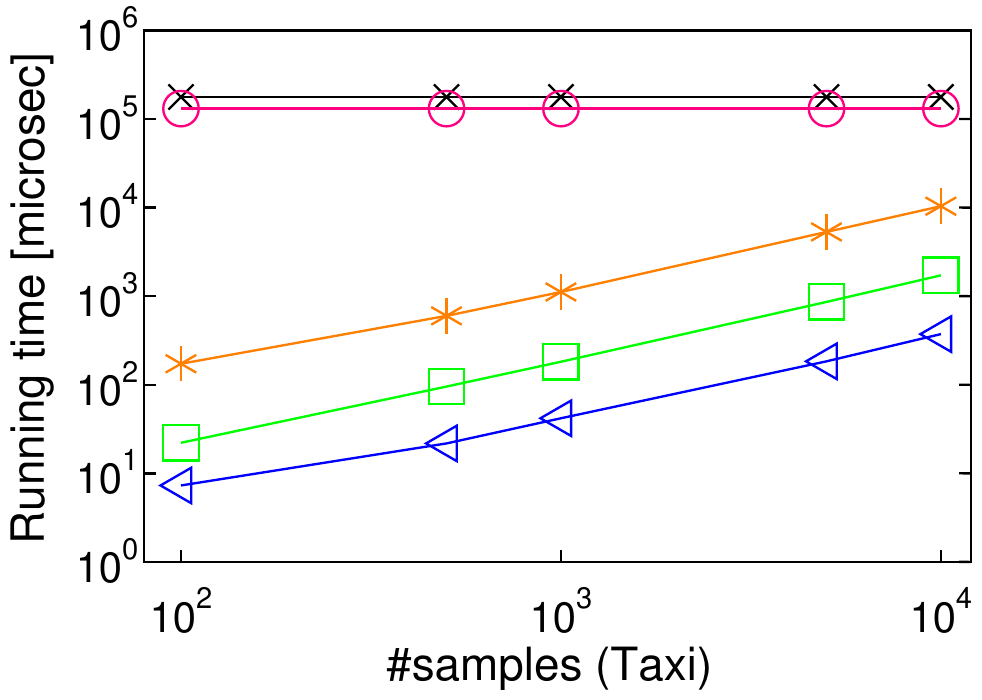}      \label{fig:sample_taxi}}
        \caption{Impact of sample size in non-weighted case.
        $\times$ shows Interval tree, \textcolor{magenta}{$\circ$} shows \textcolor{magenta}{\hint}, \textcolor{orange}{$\ast$} shows \textcolor{orange}{KDS}, \textcolor{blue}{$\triangleleft$} shows \textcolor{blue}{AIT}, and \textcolor{green}{$\square$} shows \textcolor{green}{AIT-V}.}
        \label{fig:sample}
    \end{center}
\end{figure}

\vs
\noindent
\textbf{Impact of sample size.}
Next, we investigate the impact of sample size $s$, which is shown in Fig. \ref{fig:sample}.
Interval tree and \hint are not affected by $s$, since their dominant costs are derived from computing $q \cap X$, as shown in Tables \ref{tab:candidate-time} and \ref{tab:sampling-time}.
As with our theoretical analysis, the running times of our algorithms (and KDS) are linear to $s$.
Even when many samples (e.g., $s = 10,000$) are required, our algorithms are still faster than the others, while the performance of KDS can be worse than the search-based algorithms.

\begin{figure}[!t]
    \begin{center}
        \subfigure[Book]{%
    	\includegraphics[width=0.47\linewidth]{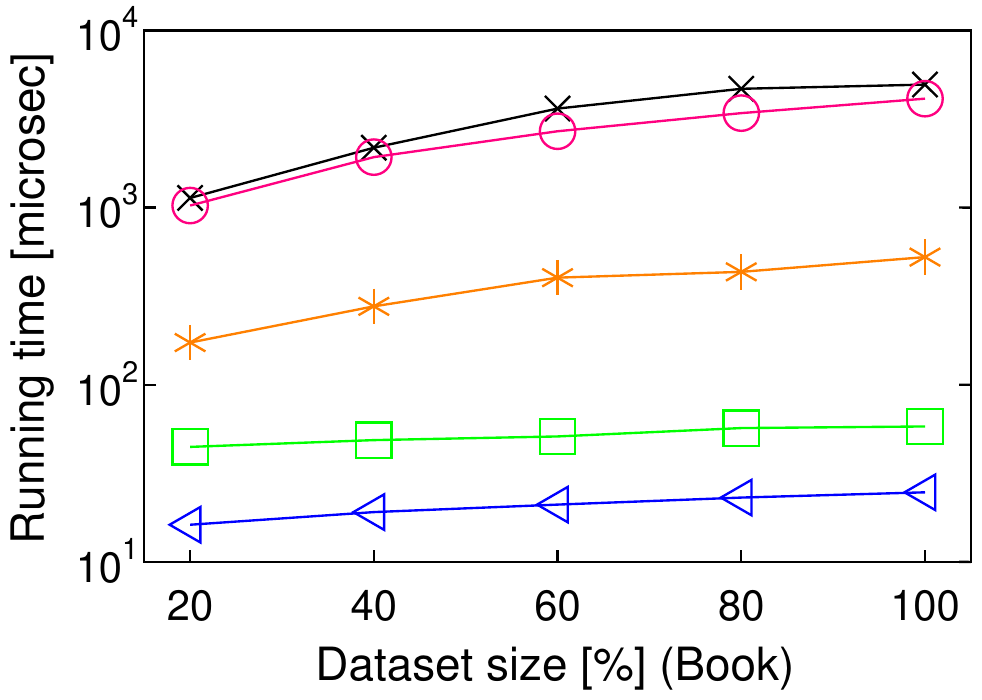}     \label{fig:cardinality_book}}
        \subfigure[BTC]{%
    	\includegraphics[width=0.47\linewidth]{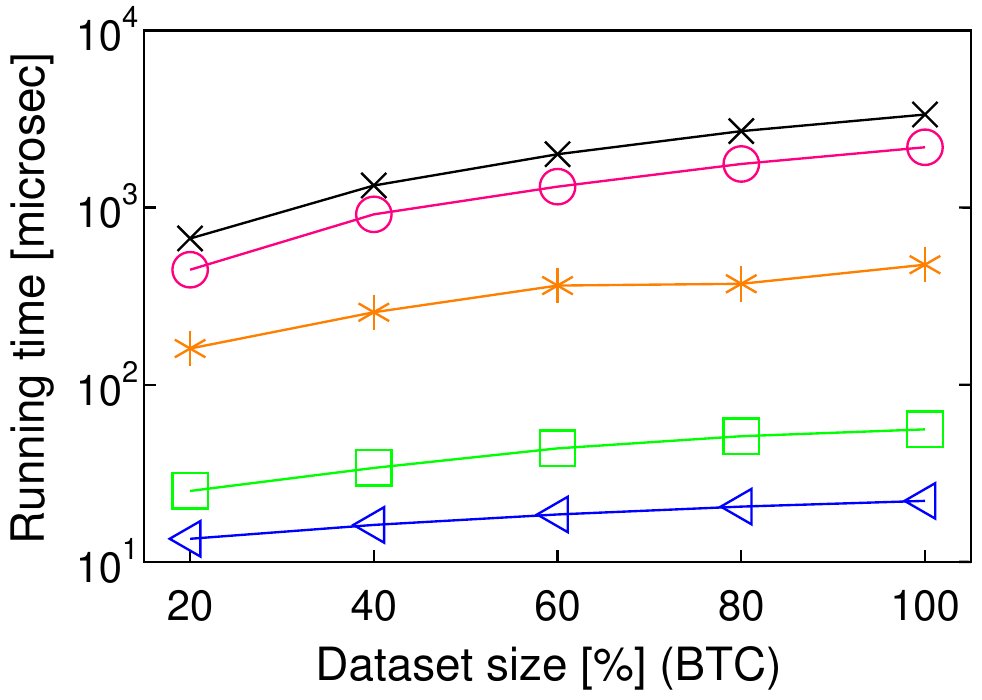}      \label{fig:cardinality_btc}}
        \subfigure[Renfe]{%
            \includegraphics[width=0.47\linewidth]{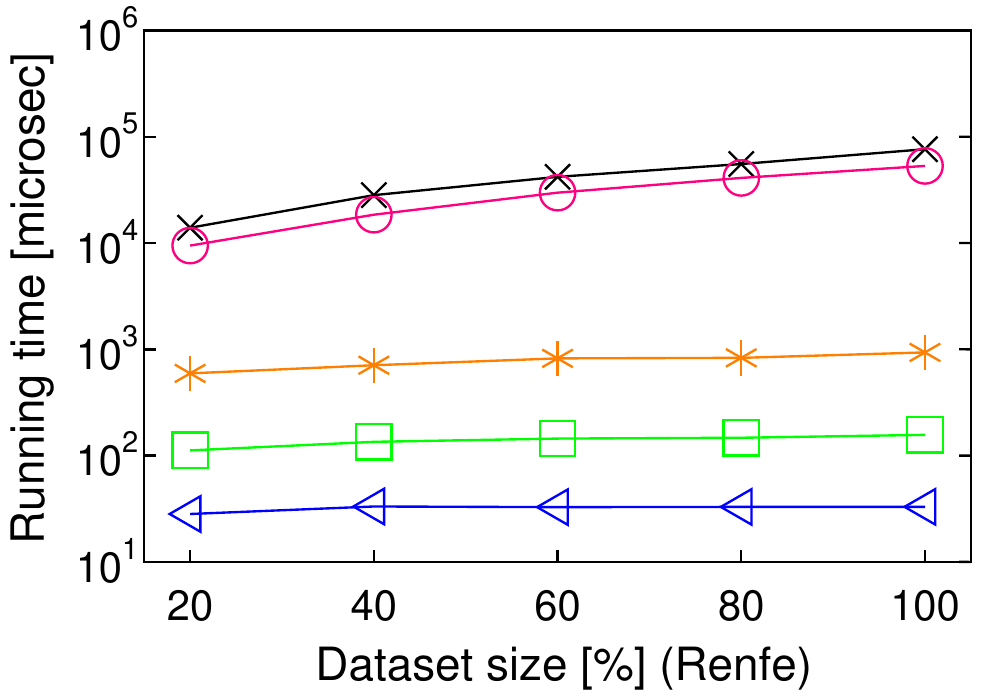}     \label{fig:cardinality_renfe}}
        \subfigure[Taxi]{%
    	\includegraphics[width=0.47\linewidth]{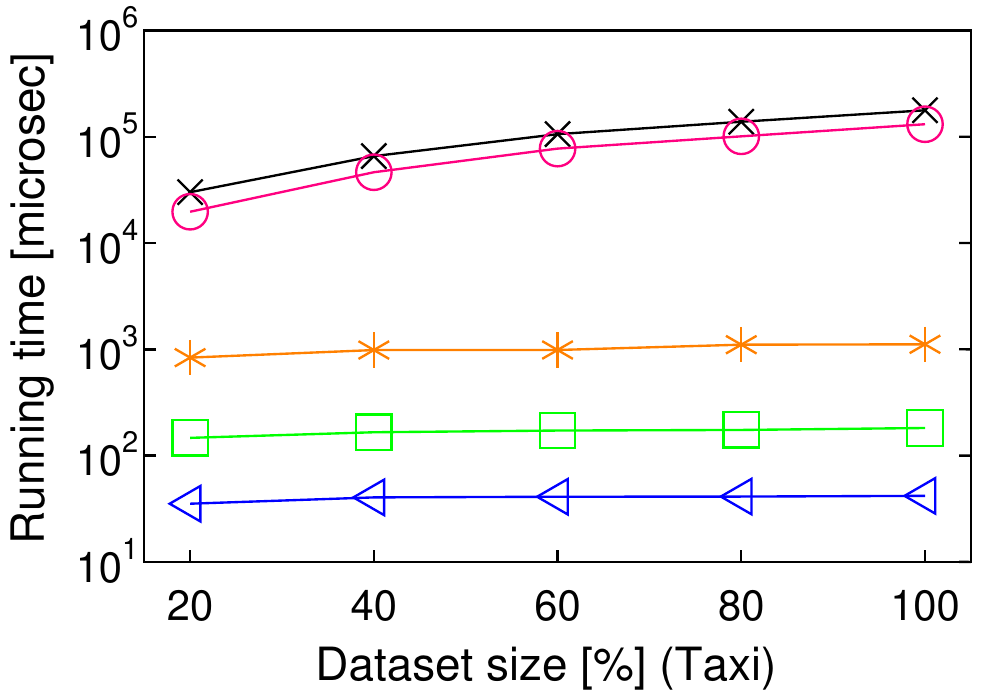}     \label{fig:cardinality_taxi}}
        \caption{Impact of dataset size in non-weighted case.
        $\times$ shows Interval tree, \textcolor{magenta}{$\circ$} shows \textcolor{magenta}{\hint}, \textcolor{orange}{$\ast$} shows \textcolor{orange}{KDS}, \textcolor{blue}{$\triangleleft$} shows \textcolor{blue}{AIT}, and \textcolor{green}{$\square$} shows \textcolor{green}{AIT-V}.}
        \label{fig:cardinality}
    \end{center}
\end{figure}

\vs
\noindent
\textbf{Impact of dataset size.}
Fig. \ref{fig:cardinality} reports the scalability to the dataset size.
The results of Interval tree and \hint suggest that $|q \cap X| = \Omega(n)$.
Our algorithms need only tens of microseconds and are clearly not sensitive to the dataset size.
The importance of designing an $\tilde{O}(s)$ time algorithm can be seen from this result.

\vs
\noindent
\textbf{Update time.}
We introduce the results of AIT update experiments.
For the insertion test, we first built an AIT on $n - 5000$ intervals and inserted the remaining 5000 intervals into the AIT.
For the deletion test, we built an AIT on $X$ and then removed 5000 intervals from the AIT.
Table \ref{tab:update} shows the amortized update time per insertion/deletion.
One-by-one insertion incurs a long update time even for a single insertion, whereas the batch insertion substantially reduces the update time.
Also, a deletion can be handled quickly, suggesting the usefulness of the traversal approach in Algorithm \ref{algo:uniform}.

\begin{table}[!t]
    \centering
    \caption{Amortized update time of AIT [millisec]}
    \label{tab:update}
     \begin{tabular}{lrrrr} \toprule
        Dataset         & Book      & BTC       & Renfe     & Taxi      \\ \midrule
        Insertion       & 448.18    & 894.44    & 2283.23   & 6312.70   \\ 
        Batch insertion & 3.01      & 2.14      & 5.25      & 10.43     \\
        Deletion        & 2.23      & 3.24      & 31.58     & 90.38     \\ \bottomrule
    \end{tabular}
\end{table}

\subsection{Result on Weighted Intervals}
\noindent
\textbf{Pre-processing time and memory usage.}
Tables \ref{tab:preprocessing_weighted} shows the pre-processing times and memory usages of AWIT.
As with the results on non-weighted intervals, AWIT does not consume long pre-processing times, and only a little additional cost is incurred.
This is also the case for the memory consumption of AWIT.

\begin{table}[!t]
    \centering
    \caption{Pre-processing time [sec] and memory usage [GB] of AWIT}
    \label{tab:preprocessing_weighted}
    \begin{tabular}{lrrrr} \toprule
        Dataset             & Book  & BTC   & Renfe     & Taxi      \\ \midrule
        Pre-processing time & 3.15  & 6.07  & 109.86    & 282.81    \\
        Memory usage        & 0.44  & 1.13  & 12.29     & 46.15     \\ \bottomrule
    \end{tabular}
\end{table}

\vs
\noindent
\textbf{Comparison with Interval tree, \hint, and KDS.}
We next study the query processing time, and Table \ref{tab:weighted_sampling-time} shows the average time to sample $s$ intervals.
Note that the candidate computation times are the same as in Table \ref{tab:candidate-time}.
Interval tree and \hint incur substantial sampling times, different from the results in Table \ref{tab:sampling-time}.
To correctly enable weighted random sampling, they need to build an alias after $q \cap X$ is obtained.
Building it requires $O(|q \cap X|)$ time, so we have this result.

On the other hand, AWIT does not have such a drawback, and they defeat the search-based algorithms and KDS w.r.t. both candidate computation and sampling times.
The sampling time of AWIT is longer than those in the non-weighted case.
This observation is derived from the $O(\log n)$ factor in sampling a weighted random interval.

\begin{table}[!t]
    \centering
    \caption{Sampling time [microsec] (weighted case).
    Alias building time is included.}
    \label{tab:weighted_sampling-time}
     \begin{tabular}{lrrrr} \toprule
        Dataset                 & Book      & BTC       & Renfe     & Taxi      \\ \midrule
        Interval tree \& \hint  & 6594.67   & 6593.22   & 122169.91 & 389509.09 \\
        KDS                     & 1307.50   & 1442.94   & 1917.36   & 2101.71   \\
        AWIT                    & 136.39    & 134.06    & 347.94    & 446.72    \\ \bottomrule
    \end{tabular}
\end{table}

\begin{figure}[!t]
    \begin{center}
        \subfigure[Book]{%
    	\includegraphics[width=0.47\linewidth]{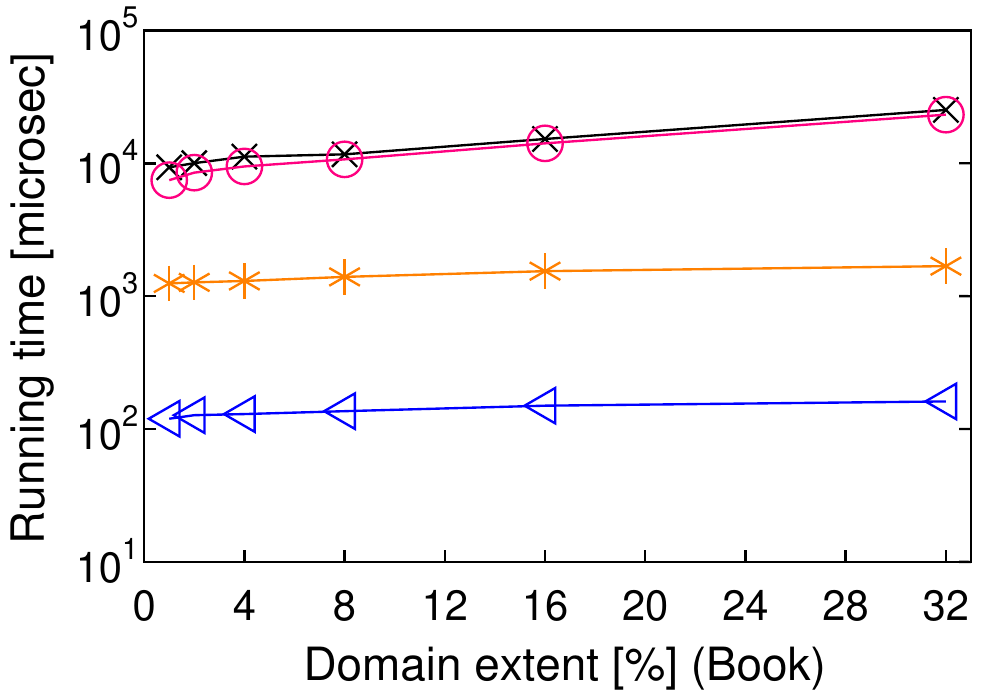}     \label{fig:extent_weighted_book}}
        \subfigure[BTC]{%
    	\includegraphics[width=0.47\linewidth]{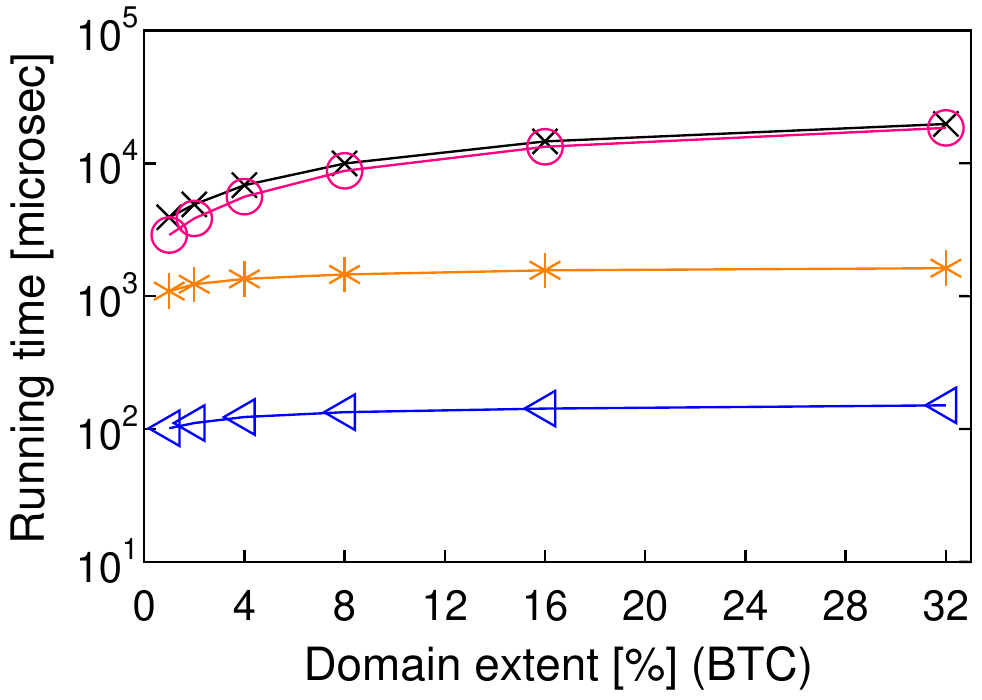}      \label{fig:extent_weighted_btc}}
        \subfigure[Renfe]{%
            \includegraphics[width=0.47\linewidth]{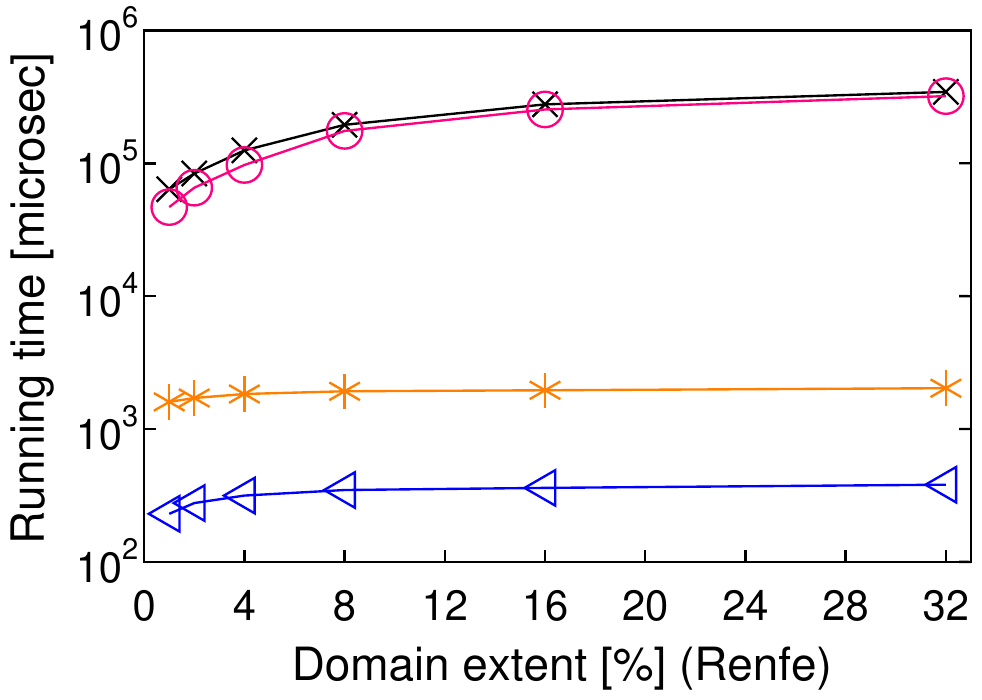}    \label{fig:extent_weighted_renfe}}
        \subfigure[Taxi]{%
    	\includegraphics[width=0.47\linewidth]{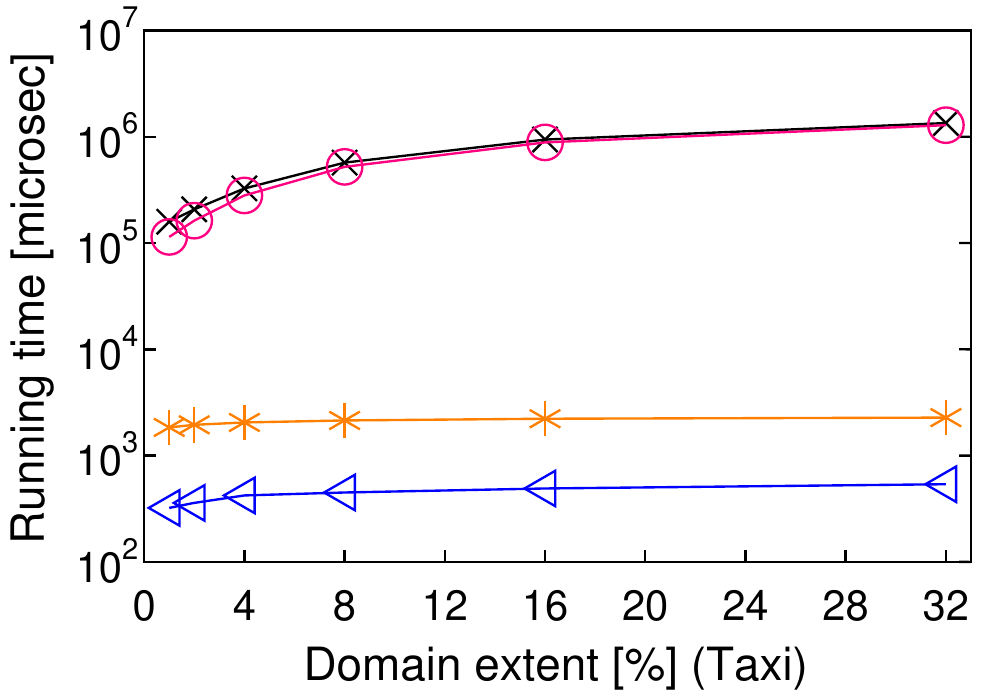}     \label{fig:extent_weighted_taxi}}
        \caption{Impact of domain extent (or query interval length) in weighted case.
        $\times$ shows Interval tree, \textcolor{magenta}{$\circ$} shows \textcolor{magenta}{\hint}, \textcolor{orange}{$\ast$} shows \textcolor{orange}{KDS}, and \textcolor{blue}{$\triangleleft$} shows \textcolor{blue}{AWIT}.}
        \label{fig:extent_weighted}
    \end{center}
\end{figure}

\vs
\noindent
\textbf{Impact of query interval length.}
Fig. \ref{fig:extent_weighted} shows the impact of query interval length (domain extent).
Although it is slight, our algorithms are affected by query interval length (e.g., see Fig. \ref{fig:extent_weighted_taxi}).
This is derived from the cumulative sum method.
As large query interval lengths yield large $\textsf{idx}^{r} - \textsf{idx}^{l}$, the binary search cost in the cumulative sum method slightly increases.

\begin{figure}[!t]
    \begin{center}
        \subfigure[Book]{%
    	\includegraphics[width=0.47\linewidth]{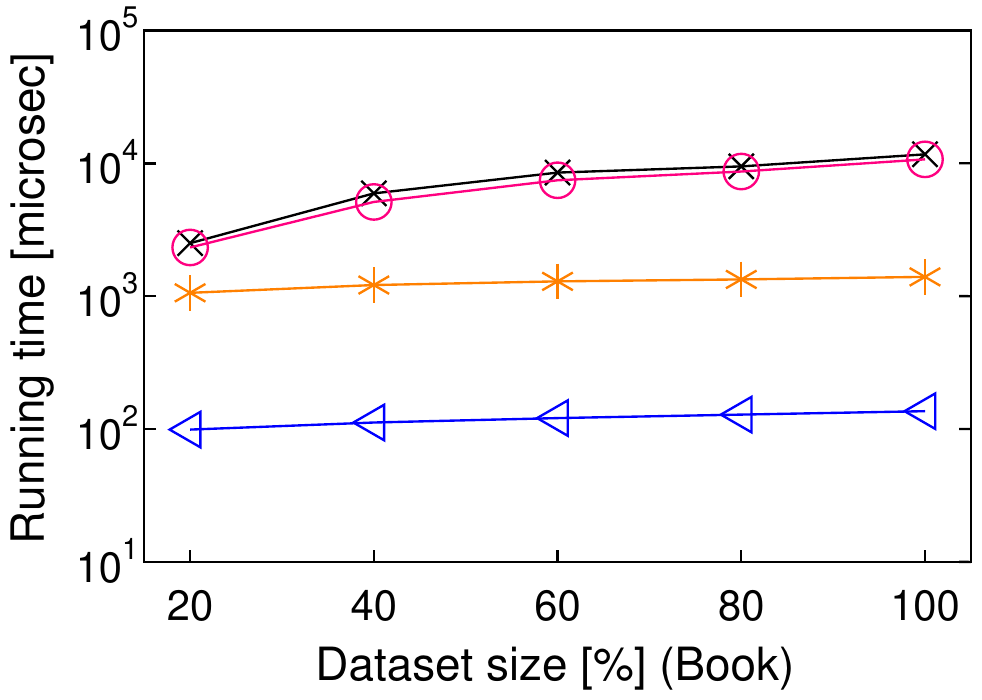}     \label{fig:cardinality_weighted_book}}
        \subfigure[BTC]{%
    	\includegraphics[width=0.47\linewidth]{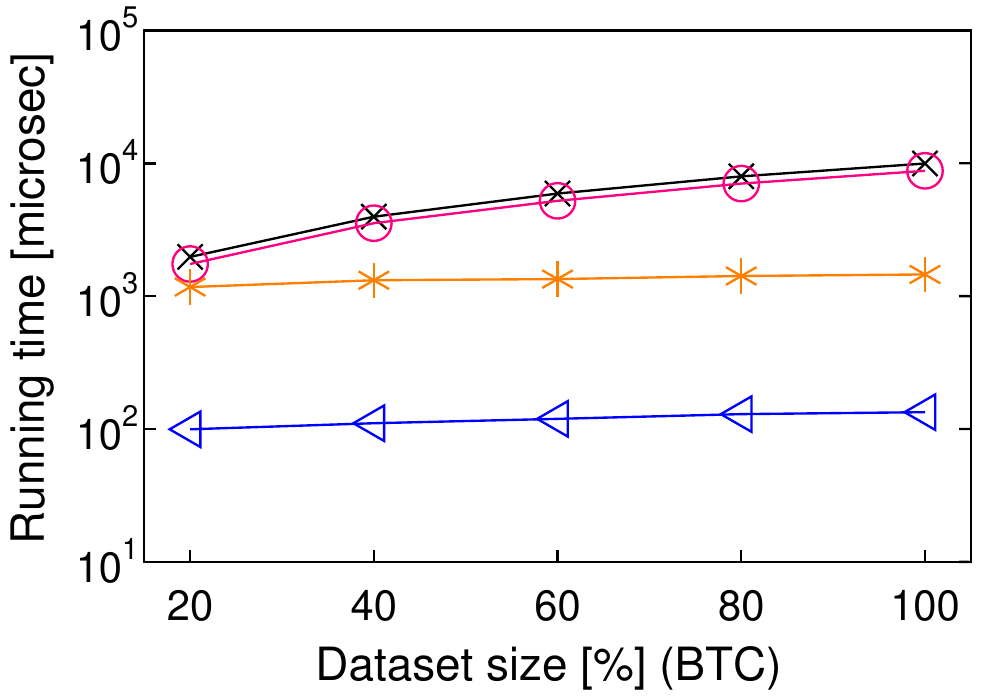}      \label{fig:cardinality_weighted_btc}}
        \subfigure[Renfe]{%
            \includegraphics[width=0.47\linewidth]{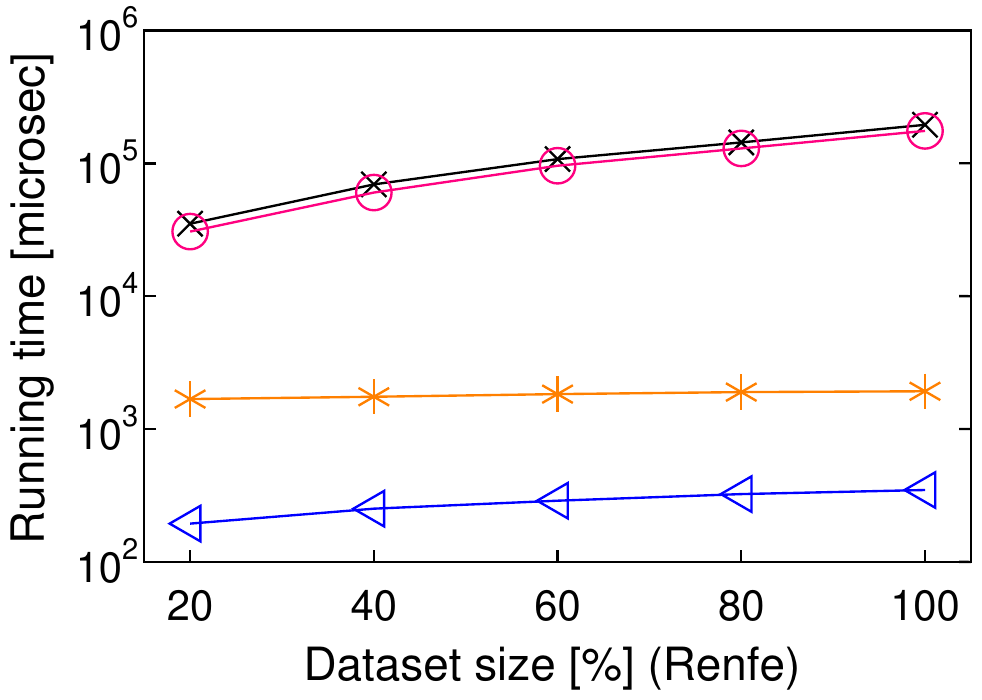}    \label{fig:cardinality_weighted_renfe}}
        \subfigure[Taxi]{%
    	\includegraphics[width=0.47\linewidth]{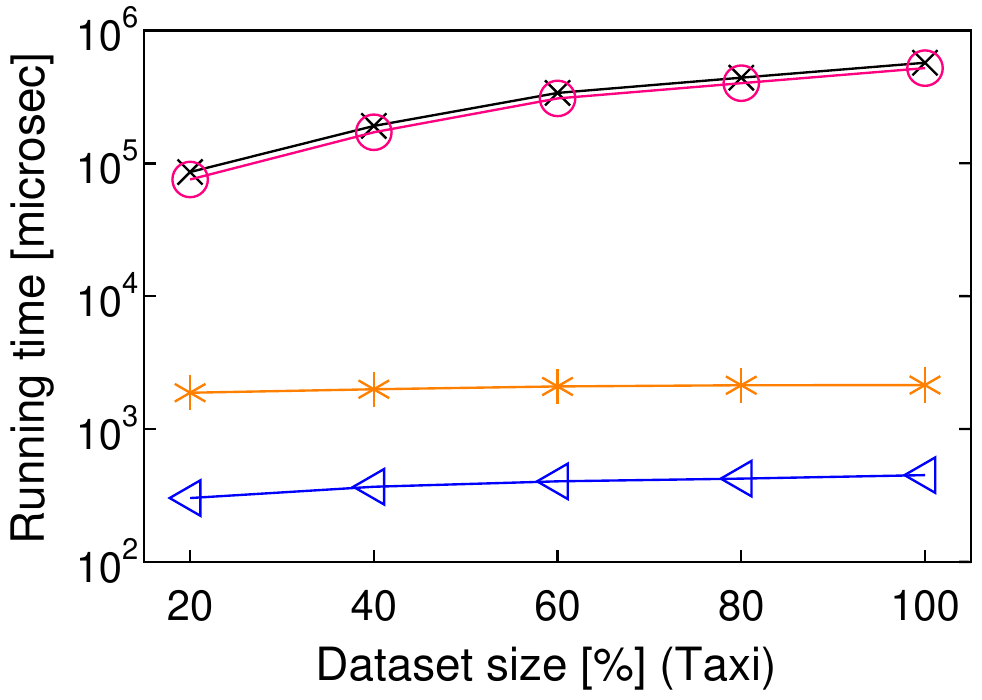}     \label{fig:cardinality_weighted_taxi}}
        \caption{Impact of dataset size in weighted case.
        $\times$ shows Interval tree, \textcolor{magenta}{$\circ$} shows \textcolor{magenta}{\hint}, \textcolor{orange}{$\ast$} shows \textcolor{orange}{KDS}, and \textcolor{blue}{$\triangleleft$} shows \textcolor{blue}{AWIT}.}
        \label{fig:cardinality_weighted}
    \end{center}
\end{figure}

\vs
\noindent
\textbf{Impact of dataset size.}
We show the scalability of each algorithm in Fig. \ref{fig:cardinality_weighted}.
Our algorithms need $\tilde{O}(s)$ time, so they again are not sensitive to the dataset size, as with the results in Fig. \ref{fig:cardinality}.
We omit the impact of sample size, because it essentially yields the same result as that in Fig. \ref{fig:sample} and the page space is limited.

\subsection{Range Counting Result}
Last, we show the range counting performance of AIT, and we compare it with $k$d-tree, which runs a range counting query in $O(\sqrt{n})$ time, and a counting version of \hint.
Table \ref{tab:range-counting} exhibits the result.
AIT is much faster than $k$d-tree and \hint, which demonstrates the superiority of AIT even in the range counting problem (a different problem to IRS).

\begin{table}[!t]
    \centering
    \caption{Range counting time [microsec]}
    \label{tab:range-counting}
    \begin{tabular}{lrrrr} \toprule
        Dataset     & Book  & BTC   & Renfe     & Taxi  \\ \midrule
        AIT         & 0.91  & 0.75  & 1.40      & 1.66  \\
        \hint       & 46.60 & 51.05 & 1156.20   & 3276.87   \\
        $k$d-tree   & 83.55 & 12.51 & 7.09      & 41.02 \\ \bottomrule
    \end{tabular}
\end{table}

\section{Related Work}  \label{sec:related_work}
\noindent
\textbf{Range search on interval data.}
Due to the importance of the interval search problem, data structures for efficient search have been developed.
Perhaps, the most famous data structure is interval tree \cite{edelsbrunner1980dynamic}, and we used this structure as our building block.
In the background of temporal databases, timeline index \cite{kaufmann2013timeline} was proposed, and it is implemented in SAP-HANA \cite{farber2012sap}.
Period index \cite{behrend2019period} was also devised in the temporal database background.
Recently, Christodoulou et al. proposed \hint \cite{christodoulou2022hint}, a hierarchical interval index for range queries.
(It is a heuristic algorithm and has no theoretical time bound for range queries.)
This structure is designed so that it can (i) achieve comparison-free, (ii) easily adapt to the distribution of a given dataset (e.g., sparsity), and (iii) exploit hardware-aware optimization.
In \cite{christodoulou2022hint,christodoulou2023hint}, it is empirically shown that \hint outperforms the interval tree, the timeline index, and the period index.
We therefore used \hint as a competitor, and Section \ref{sec:experiment} clarifies that even the state-of-the-art range search algorithm cannot efficiently solve the IRS problem.

\vs
\noindent
\textbf{Other operations on interval data.}
Segment-tree \cite{mark2008computational} is also a famous data structure.
It needs $O(n\log n)$ space and supports $O(\log n + K)$ time stabbing query, where $K$ is the output size.
(This structure does not support efficient range search.)
For weighted intervals, literatures \cite{agarwal2005optimal,rahul2014general,xu2017efficiently,amagata2024efficient,amagata2024efficient_} addressed the problem of finding $k$ intervals with the largest weight among the intervals stabbing a given query.
The interval join problem, which finds all pairs of intervals overlapping each other, has also been studied \cite{bouros2021memory,piatov2021cache,piatov2016interval,bouros2017forward,dignos2014overlap,bouros2020band,bouros2018interval}.
Given a set $X$ of intervals, literature \cite{cafagna2017disjoint} tackled the problem of disjoint partitioning $X$ for efficient interval join and aggregation.
Because the objective of this partitioning is different from ours in Section \ref{sec:uniform:aitv}, the technique proposed in \cite{cafagna2017disjoint} is not available for our problem.

\vs
\noindent
\textbf{IRS on other data.}
The IRS problem has been extensively studied, and its usefulness has been recognized since the 1980s \cite{olken1989random}.
The importance of ``independent samples'' has spread from \cite{hu2014independent}, which solved the IRS problem on dynamic one-dimensional data and external-memory-resident data.
Its weighted version has recently been addressed in \cite{zhange2023efficient}.
Recall that Section \ref{sec:introduction} has clarified the IRS algorithm for one-dimensional data cannot correctly solve our problem.

The IRS problem on multi-dimensional data has been addressed as well.
Afshani and Wei \cite{afshani2017independent} considered two- and three-dimensional points, whereas Afshani and Phillips \cite{afshani2019independent} assumed three-dimensional weighted points.
These works provided theoretical results for specific queries (e.g., three-dimensional half-space queries \cite{afshani2019independent}), which are different from range queries on interval data.
As a practically efficient algorithm, Xie et al. proposed KDS \cite{xie2021spatial}, which solves the IRS problem on $d$-dimensional Euclidean points in $O(n^{1-\frac{1}{d}} + s)$ expected time.
Since intervals can be mapped to a two-dimensional Euclidean space and query intervals can be mapped to orthogonal ranges \cite{rahul2014general}, we used KDS as a competitor and demonstrated that our interval-specific data structures provide much faster candidate computation and sampling than KDS theoretically (see Table \ref{tab:complexity}) and empirically (cf. Section \ref{sec:experiment}).

The above works considered low-dimensional points, and recent works \cite{aumuller2022sampling,aumuller2020fair,har2019near,aumuller2021fair,aumuller2022sampling_} addressed the IRS problem on high-dimensional spaces.
These works consider sampling a single random object with a distance at most a threshold from a given query.
They show that such sampling can be done in sub-linear (expected) time to $n$ with high probability \cite{aumuller2020fair} or approximate independence probabilities \cite{har2019near}.
This setting is totally different from ours, and our theoretical results do not have this probability condition and approximation.

\section{Conclusion}    \label{sec:conclusion}
We addressed the problems of independent range sampling on non-weighted and weighted interval data.
To theoretically and practically solve these problems efficiently, we proposed some variants of interval tree and $\tilde{O}(s)$ time algorithms, where $s$ is the number of samples.
We demonstrated that, for sufficiently large datasets, these algorithms are theoretically faster than existing techniques that can be used for the problems.
In addition, from extensive experiments, we confirmed that our algorithms outperform competitors by large margins.

\section*{Acknowledgements}
This work was partially supported by AIP Acceleration Research JPMJCR23U2 and Adopting Sustainable Partnerships for Innovative Research Ecosystem JPMJAP2328, JST.

\balance
\bibliographystyle{IEEEtran}
\bibliography{sample}

\end{document}